\documentclass[letterpaper,11pt]{article}
\pdfoutput=1 

\usepackage{jheppub} 

\usepackage[T1]{fontenc} 
\usepackage{lmodern}
\usepackage[mathscr]{euscript}
\usepackage{graphicx}
\usepackage{ytableau}
\usepackage{amsmath}
\usepackage{amssymb}
\usepackage{hyperref}
\usepackage{xcolor}
\usepackage{color}
\usepackage{dsfont}

\DeclareFontFamily{U}{mathx}{\hyphenchar\font45}
\DeclareFontShape{U}{mathx}{m}{n}{
      <5> <6> <7> <8> <9> <10>
      <10.95> <12> <14.4> <17.28> <20.74> <24.88>
      mathx10
      }{}
\DeclareSymbolFont{mathx}{U}{mathx}{m}{n}
\DeclareFontSubstitution{U}{mathx}{m}{n}
\DeclareMathSymbol{\bigplus}        {1}{mathx}{"90}
\DeclareMathSymbol{\bigtimes}       {1}{mathx}{"91}

\DeclareFontFamily{OT1}{pzc}{}
\DeclareFontShape{OT1}{pzc}{m}{it}{<-> s * [1.10] pzcmi7t}{}
\DeclareMathAlphabet{\mathpzc}{OT1}{pzc}{m}{it}
\newenvironment{rcases}
  {\left.\begin{aligned}}
  {\end{aligned}\right\rbrace}

\title{\boldmath{Splitting of surface defect partition functions and integrable systems}}

\author[a]{Saebyeok Jeong}
\preprint{YITP-SB-17-35}

\affiliation[a]{C. N. Yang Institute for Theoretical Physics, Stony Brook University,\\Stony Brook, NY 11794-3840, USA}

\emailAdd{saebyeok.jeong@gmail.com}

\abstract{We study Bethe/gauge correspondence at the special locus of Coulomb moduli where the integrable system exhibits the splitting of degenerate levels. For this investigation, we consider the four-dimensional pure $\EuScript{N}=2$ supersymmetric $U(N)$ gauge theory, with a half-BPS surface defect constructed with the help of an orbifold or a degenerate gauge vertex. We show that the non-perturbative Dyson-Schwinger equations imply the Schr\"{o}dinger-type and the Baxter-type differential equations satisfied by the respective surface defect partition functions. At the special locus of Coulomb moduli the surface defect partition function splits into parts. We recover the Bethe/gauge dictionary for each summand.}


\begin{document} 
\maketitle
\flushbottom

\section{Introduction}
\label{sec:intro}
Supersymmetric gauge theories in various dimensions exhibit diverse connections with integrable systems. The four-dimensional gauge theory with $\EuScript{N}=2$ supersymmetry is one of the interesting cases to consider. The common feature of this class of theories is that the low-energy description achieved in \cite{sw1,sw2} naturally reveals the structure of an algebraic integrable system \cite{swint1,swint2}. The correspondence was promoted to the quantum level in \cite{neksha3}, by putting the $\EuScript{N}=2$ gauge theory into the general framework of the Bethe/gauge correspondence \cite{neksha1,neksha2}. When subject to the Nekrasov-Shatashvili limit of the $\Omega$-deformation $(\varepsilon_1 =\hbar, \varepsilon_2 \rightarrow 0)$, the four-dimensional $\EuScript{N}=2$ gauge theory effectively becomes a two-dimensional theory with $\EuScript{N}=(2,2)$ supersymmetry. The general Bethe/gauge correspondence states the chiral ring is the set of quantum Hamiltonians, while the set of supersymmetric vacua is identified with the (Hilbert) space of the corresponding quantum integrable system,
\begin{align}
\vert \text{eigen} \rangle \longleftrightarrow \text{vac}.
\end{align}
In particular, the spectrum of the Hamiltonian is calculated as the gauge theory vacuum expectation value of the corresponding chiral observable in the Nekrasov-Shatashvili limit,
\begin{align}
\langle \text{eigen} \vert H_\EuScript{O} \vert \text{eigen} \rangle = \langle \EuScript{O} \rangle \vert_{\varepsilon_2 \rightarrow 0,  \text{vac}}. \label{spectrum}
\end{align}
The chiral ring is spanned by the gauge-invariant observables $\EuScript{O}_k = \text{Tr} \phi^k$, where $\phi$ is the complex scalar in the $\EuScript{N}=2$ vector multiplet. In generic case, their vacuum expectation values are finite in the Nekrasov-Shatashvili limit, since they reduce to the vacuum expectation values of the twisted chiral observables in the effective two-dimensional $\EuScript{N}=(2,2)$ theory. Therefore the right hand side of the dictionary \eqref{spectrum} is well-defined, providing a way to compute the spectrum of the quantum Hamiltonian from gauge theory perspective. Note that the partition function of the gauge theory shows the asymptotic behavior $\text{log} \EuScript{Z} = \frac{\widetilde{\EuScript{W}}}{\varepsilon_2} + \mathcal{O} (\varepsilon_2 ^0)$ in the Nekrasov-Shatashvili limit, where $\widetilde{\EuScript{W}}$ is the effective twisted superpotential of the effective two-dimensional theory.

The equations which describe the vacua in the low-energy theory correspond to the quantization conditions on the integrable system side. The Nekrasov-Shatashvili limit of the four-dimensional $\EuScript{N}=2$ gauge theory leads to several inequivalent quantization schemes, in particular, the type A and the type B quantizations \cite{neksha3,nekwit}. In the present work we mainly focus on the type B quantization, in which we impose the condition
\begin{align}
\text{exp} \left( 2 \pi i \frac{a_\alpha}{\varepsilon_1} - i \theta_\alpha \right) =1, \quad \theta_\alpha \in [0, 2\pi). \label{typeb}
\end{align} 
Note that the $\theta$-angles can be introduced in a gauge-invariant fashion. Namely, for given values of the gauge-invariant coordinates on the Coulomb moduli space, $\langle \EuScript{O}_k \rangle = \langle \text{Tr} \phi^k \rangle$, the Coulomb moduli $a_\alpha$ are determined up to the permutations with each other. Therefore in the real slice that we are choosing in the type B quantization, $\frac{a_\alpha}{\varepsilon_1} \in \mathbb{R}$, the $\theta$-angles are determined up to the permutations with each other. For the quantization condition \eqref{typeb}, we look for the eigenfunctions which are quasi-periodic with the Bloch angles ($\theta_\alpha$). For example, for the pure $\EuScript{N}=2$ theory and for the $\EuScript{N}=2^*$ theory with the gauge group $U(N)$, the formula \eqref{spectrum} under the condition \eqref{typeb} computes the spectrum of the Hamiltonians of the $N$-particle periodic Toda system and the $N$-particle elliptic Calogero-Moser system respectively, whose eigenfunctions are quasi-periodic with the Bloch angles ($\theta_\alpha$). Note that the spectrum would have been ($N!$)-fold degenerate in the non-interacting limit had we tuned all the Bloch angles to be the same. For generic values of Bloch angles, the $S_N$-symmetry of the 0-th order wavefunctions is completely broken, leaving non-degenerate level for each spectrum.

We can revive some of the degenerate levels at the 0-th order by tuning the corresponding Bloch angles, e.g. as $\theta_\alpha = \theta_\beta$. The integrable system is still well-defined, and the eigenvalues are expected to be non-degenerate. However, we observe the missing link in the correspondence with the gauge theory. According to the condition \eqref{typeb}, tuning the Bloch angles as $\theta_\alpha = \theta_\beta$ is equivalent to investigaing the special locus of Coulomb moduli, $\left\{ \frac{a_{\alpha \beta}}{\varepsilon_1} \in \mathbb{Z} \setminus \{0\} \right\}$.\footnote{We have excluded $a_{\alpha \beta} =0$ since in this case the splitting of the degeneracy at the 0-th order does not occur and \eqref{spectrum} works as it is.} At the locus, the formula \eqref{spectrum} breaks down since the right hand side becomes divergent due to the additional singularities in $\varepsilon_2 \rightarrow 0$. The asymptotic behavior of the partition function is no longer $\text{log} \EuScript{Z} = \frac{\widetilde{\EuScript{W}}}{\varepsilon_2} + \mathcal{O}(\varepsilon_2 ^0)$, and the effective twisted superpotential cannot be properly obtained by just taking $\widetilde{\EuScript{W}} = \lim_{\varepsilon_2 \to 0} \varepsilon_2 \text{log} \EuScript{Z}$. Inspired by the well-established correspondence for the generic value of the Coulomb moduli, we now may attempt to recover the correspondence at the special locus, especially by first investigating the perturbative series in the integrable system side. This is the main subject of the present work.

We may try to approach the special locus of Coulomb moduli from the gauge theory with partial $\Omega$-deformation and partial noncommutativity. Instead of turning on both $\Omega$-deformation parameters and then taking the Nekrasov-Shatashvili limit, we can from the beginning turn on one of the parameters $\varepsilon_1$ only. When the noncommutativity along the $\varepsilon_1$-plane is turned on, the four-dimensional $\EuScript{N}=2$ theory can be described by a two-dimensional $\EuScript{N}=(2,2)$ theory with an infinite dimensional gauge group. The investigation shows that the only massless modes around the trivial vacuum are the diagonal components of the gauge multiplet, which is consistent with the expectation that the low-energy effective theory is in Coulomb phase without any matter. However, when the Coulomb moduli assume the special values as $\frac{a_{\alpha \beta}}{\varepsilon_1} \in \mathbb{Z} \setminus \{0\} $, additional massless matter multiplets seem to arise, signifying the failure of the effective description.

The surface defect provides a tool for the investigation. The four-dimensional gauge theory with a half-BPS surface defect can be viewed as the theory on an orbifold. The equivariant localization computation applied for the bulk theory immediately generalizes to compute the surface defect partition function \cite{kantachi}. The gauge theory observables are also naturally generalized to the theory in the presence of the surface defect. In particular, an important class of observables, called the $qq$-character, has its fractionalized counterpart in the theory with the surface defect \cite{nek2}. In \cite{nek4,nek7} the $qq$-characters with and without the surface defect were realized as the orbifolded crossed instanton partition functions. The compactness theorem proved in \cite{nek3} implied a certain vanishing theorem for the expectation value of the $qq$-characters. The vanishing equations, called the non-perturbative Dyson-Schwinger equations, can be used to derive the KZ equation satisfied by the surface defect partition function of quiver gauge theory \cite{nek8}. In this paper, we show that the Dyson-Schwinger equation in the presence of the surface defect produces a Schr\"{o}dinger-type equation satisfied by the orbifold surface defect partition function of the pure $U(N)$ gauge theory. Therefore the surface defect partition function provides a constructive approach to the eigenstate wavefunctions as well as the spectra of the Hamiltonians of the corresponding quantum integrable system. 

Another type of half-BPS surface defect in quiver gauge theories can be obtained by a 2d-4d combined system \cite{agt2,gai2}. In this construction, a half-BPS surface defect is constructed by weakly coupling the flavor group of the two-dimesional gauged linear sigma model to the four-dimensional gauge field. In the IR, the effective action gets the twisted F-term contribution from the two-dimensional sigma model which has the Higgs branch as its target. This type of surface defect was identified with the insertion of a simplest fully-degenerate field in the Toda CFT side \cite{agt2}. In the dictionary of \cite{agt,agtw}, this is equivalent to tuning the mass of a fundamental hypermultiplet coupled to a vector multiplet, in such a way that the contributions to the instanton partition function only come from single-column Young diagrams (or single-row, depending on which plane we put the surface defect) \cite{nek7}. For brevity, let us call a gauge vertex \textit{degenerate} for such cases. The Dyson-Schwinger equations can be used to derive the BPZ equation satisfied by the partition function of the quiver gauge theory with a degenerate gauge vertex \cite{nek8}. In this paper, we show that the pure $U(N)$ gauge theory partition function with this type of surface defect satisfies a Baxter-type equation.

The main observation of this work is that the orbifold surface defect partition function at the special locus $ \left\{\frac{a_{\alpha \beta}}{\varepsilon_1} \in \mathbb{Z} \setminus \{0\} \right\}$ splits into parts, schematically,
\begin{align}
\boldsymbol{\Psi} = \sum_{\gamma} \boldsymbol{\Psi}_\gamma.
\end{align} 
This behavior accounts for the level splitting on the integrable system side. Each part of the surface defect partition function shows the proper asymptotic behavior of $\text{log} \boldsymbol{\Psi}_\gamma = \frac{\widetilde{\EuScript{W}}_\gamma}{\varepsilon_2} + \mathcal{O}(\varepsilon_2 ^0)$, and the dictionary \eqref{spectrum} is recovered to reproduce the spectrum of each split level. It should be noted that each split part $\boldsymbol{\Psi}_\gamma$ of the surface defect partition function shows the series expansion in fractional powers of the gauge coupling, which correctly accounts for the series expansions of the spectra of the split levels.

The rest of the paper is organized as follows. In section \ref{generalities}, we review the instanton partition function of the four-dimensional $\EuScript{N}=2$ quiver gauge theory, along with various gauge-invariant observables including the $qq$-characters. The Bethe/gauge correspondence is then explained with a description of two inequivalent types of quantization. In section \ref{partial}, we study the special locus of Coulomb moduli in the four-dimensional gauge theory with partial $\Omega$-deformation and partial noncommutativity. The investigation reveals the emergence of additional massless modes, which indicates a failure of the effective description of the theory. In section \ref{surface}, we review the orbifold and the degenerate gauge vertex constructions of half-BPS surface defects, and compute the surface defect partition functions. We study the non-perturbative Dyson-Schwinger equations in the presence of surface defects. We verify that the partition functions of the $A_1$-theory with a regular orbifold surface defect and of the $A_2$-theory with a degenerate gauge vertex satisfy the Schr\"{o}dinger-type and the Baxter-type equations, respectively. In section \ref{splitting}, we observe that at the special locus of the Coulomb moduli, the surface defect partition function splits into parts, recovering the correspondence with the quantum integrable system. We conclude in section \ref{discussion} with possible generalizations and discussions.

\section{Generalities} \label{generalities}
Let us begin by reviewing the general facts of the quiver gauge theory and the Bethe/gauge correspondence. First we study the instanton partition function of quiver gauge theories. The well-known correspondence between the pure $\EuScript{N}=2$ supersymmetric $U(N)$ gauge theory in Nekrasov-Shatashvili limit and the quantum periodic Toda system is also discussed. We refer to \cite{nekpessha, nek2} and their references for more general discussions of the $\EuScript{N}=2$ supersymmetric quiver gauge theories.
\subsection{Instanton partition function of quiver gauge theories} \label{instpartfunc}
In this section, we generally follow the discussion in \cite{nek2}. A quiver is an oriented graph. For a given quiver $\gamma$, let $\text{Vert}_{\gamma}$ and $\text{Edge}_{\gamma}$ be the sets of vertices and edges, respectively. We define $s, t : \text{Edge}_\gamma \rightarrow \text{Vert}_\gamma$ as the maps sending an edge to its source and target, respectively. We also assign two vectors of integers,
\begin{align}
\mathbf{n} = (\mathpzc{n}_\mathbf{i})_{\mathbf{i} \in \text{Vert}_\gamma} \in \mathbb{Z}_{>0} ^{\text{Vert}_\gamma}, \quad \mathbf{m} = (\mathpzc{m}_\mathbf{i})_{\mathbf{i} \in \text{Vert}_\gamma} \in \mathbb{Z}_{\geq 0} ^{\text{Vert}_\gamma}.
\end{align}

The quiver gauge theory for $\gamma$ is the four-dimensional $\EuScript{N}=2$ supersymmetric gauge theory with the gauge group
\begin{align}
G_{g} = \bigtimes_{\mathbf{i} \in \text{Vert}_\gamma} U(\mathpzc{n}_\mathbf{i}),
\end{align}
and the flavor group
\begin{align}
G_f = \left( \bigtimes_{\mathbf{i} \in \text{Vert}_\gamma} U(\mathpzc{m}_\mathbf{i}) \times U(1) ^{\text{Edge}_\gamma} \right) \Bigg/ U(1) ^{\text{Vert}_\gamma},
\end{align}
where the overall $U(1)^{\text{Vert}_\gamma}$ gauge transformation has been quotiented out,
\begin{align}
(u_\mathbf{i})_{\mathbf{i} \in \text{Vert}_\gamma} : \left( (g_\mathbf{i})_{\mathbf{i} \in \text{Vert}_\gamma} , ( u_\mathbf{e})_{\mathbf{e} \in \text{Edge}_\gamma} \right) \mapsto \left( (u_\mathbf{i} g_\mathbf{i})_{\mathbf{i} \in \text{Vert}_\gamma} , (u_{s(\mathbf{e})} u_\mathbf{e} u_{t(\mathbf{e})} ^{-1})_{\mathbf{e} \in \text{Edge}_\gamma} \right).
\end{align}
In terms of the field content, we have the vector multiplets $\boldsymbol{\Phi} = (\Phi_\mathbf{i})_{\mathbf{i} \in \text{Vert}_\gamma}$ in the adjoint representation of $G_g$, the fundamental hypermultiplets $\boldsymbol{Q}_{\text{fund}} = (Q_\mathbf{i})_{\mathbf{i} \in \text{Vert}_\gamma}$ in the fundamental representation of $G_g$ and the antifundamental representation of $G_f$, and finally the bifundamental hypermultiplets $\boldsymbol{Q}_{\text{bifund}} = (Q_\mathbf{e})_{\mathbf{e} \in \text{Edge}_\gamma}$ in the bifundamental representation $(\overline{n_{s(\mathbf{e})}} ,n_{t(\mathbf{e})})$ of $G_g$. Given these fields, the action of the theory is fixed by the $\EuScript{N}=2$ supersymmetry up to the choice of complexified gauge couplings
\begin{align}
\mathfrak{q}_\mathbf{i} = \text{exp}(2 \pi i \tau_\mathbf{i}) \quad \left( \tau_\mathbf{i} = \frac{\vartheta_\mathbf{i}}{2\pi} + \frac{4\pi i}{g_\mathbf{i} ^2} \right), \quad \mathbf{i} \in \text{Vert}_\gamma,
\end{align}
and the masses of the hypermultiplets
\begin{align}
&\textbf{\textit{m}} = ((\mathbf{m}_\mathbf{i})_{\mathbf{i} \in \text{Vert}_\gamma}, (m_\mathbf{e})_{ \mathbf{e} \in  \text{Edge}_\gamma}), \nonumber \\
&\mathbf{m}_\mathbf{i} = \text{diag}(m_{\mathbf{i},1} , \cdots, m_{\mathbf{i},\mathpzc{m}_\mathbf{i}}) \in \text{End}(\mathbb{C}^{\mathpzc{m}_\mathbf{i}}), \quad m_\mathbf{e} \in \mathbb{C},
\end{align}
which are viewed as the equivariant parameters for the flavor group $G_f$. The vacuum expectation values of the complex scalars,
\begin{align}
\langle \Phi_\mathbf{i} \rangle = \mathbf{a}_\mathbf{i}, \quad  \mathbf{a}_\mathbf{i} = \text{diag}(a_{\mathbf{i},1}, \cdots, a_{\mathbf{i},\mathpzc{n}_\mathbf{i}}) \in \text{End}(\mathbb{C}^{\mathpzc{n}_\mathbf{i}}), \quad \mathbf{i} \in \text{Vert}_\gamma,
\end{align}
break the gauge symmetry to the maximal torus, leading to the Coulomb phase of the theory. The Coulomb moduli $\mathbf{a}= (\mathbf{a}_\mathbf{i})_{\mathbf{i} \in \text{Vert}_\gamma}$ parametrize the vacua of the theory. We also have the four-dimensional rotation group
\begin{align}
G_{\text{rot}} = SO(4),
\end{align}
for which we turn on the $\Omega$-deformations with the equivariant parameters $\boldsymbol{\varepsilon} = (\varepsilon_1, \varepsilon_2)$. We denote the total global symmetry group as
\begin{align}
H = G_g \times G_f \times G_{\text{rot}},
\end{align}
whose maximal torus is denoted by $T_H \subset H$. The instanton partition function is a $T_H$-equivariant integral over the framed noncommutative instanton moduli space, and therefore is parametrized by $(\mathbf{a}, \textbf{\textit{m}}, \boldsymbol{\varepsilon}) \in \text{Lie}(T_H)$.

The ADHM moduli space is useful for the equivariant localization, which can be explicitly written as
\begin{align}
&\EuScript{M} (n , k)  = \begin{cases} \quad \quad B_{1,2}  : K \rightarrow K, \\ I : N \rightarrow K, J : K \rightarrow N\end{cases} \Bigg\vert \begin{rcases} \left[ B_1 , B_2   \right] + I J=0, \quad \quad \quad \quad \quad   \\ [B_1 , {B_1 } ^\dagger ] + [B_2  , {B_2 } ^ \dagger] + I {I} ^\dagger - {J} ^\dagger J = \zeta \end{rcases} \Bigg/ U(k), \\
&(N = \mathbb{C}^n, K = \mathbb{C}^k) \nonumber
\end{align}
for given $n$ and $k$. Here, the parameter $\zeta$ corresponds to the noncommutativity of Euclidean spacetime. It is introduced to resolve the singularities of the Uhlenbeck compactification, on which we do not elaborate further in this paper. Given the vector of instanton charges $\mathbf{k} = (k_\mathbf{i})_{\mathbf{i} \in \text{Vert}_\gamma} \in \mathbb{Z}_{\geq 0}$, the total framed noncommutative instanton moduli space of the quiver gauge theory for $\gamma$ is
\begin{align}
\EuScript{M}_\gamma (\mathbf{n}, \mathbf{k}) \equiv \bigtimes_{\mathbf{i} \in \text{Vert}_\gamma} \EuScript{M}(\mathpzc{n}_\mathbf{i} , k_\mathbf{i}) \label{quivmodul}.
\end{align}

The $T_H$-equivariant integration over the instanton moduli space \eqref{quivmodul} localizes on the set of fixed points of $T_H$-action, $\mathcal{M}_\gamma (\mathbf{n}, \mathbf{k})^{T_H}$, which is the set of colored partitions $\boldsymbol{\lambda}=((\lambda^{(\mathbf{i}, \alpha)})_{\alpha=1} ^{\mathpzc{n}_\mathbf{i}})_{\mathbf{i} \in \text{Vert}_\gamma}$, where each $\lambda^{(\mathbf{i}, \alpha)}$ is a partition,
\begin{align} \lambda^{(\mathbf{i}, \alpha)} = \left( \lambda^{(\mathbf{i},\alpha)} _1 \geq \lambda^{(\mathbf{i}, \alpha)}_2 \geq \cdots \geq \lambda^{(\mathbf{i},\alpha)}_{l(\lambda^{(\mathbf{i},\alpha)})} > \lambda^{(\mathbf{i}, \alpha)}_{l(\lambda^{(\mathbf{i},\alpha)})+1} = \cdots = 0   \right),
\end{align}
with the size $\vert \lambda^{(\mathbf{i}, \alpha)} \vert = \sum_{i=1} ^{l(\lambda^{(\mathbf{i}, \alpha)})} \lambda^{(\mathbf{i},\alpha)} _i = k_{\mathbf{i}, \alpha}$ constrained by $k_\mathbf{i} = \sum_\alpha k_{\mathbf{i}, \alpha} = \vert \boldsymbol{\lambda} ^{(\mathbf{i})} \vert$ \cite{nek1, nekokoun}. There is an one-to-one correspondence between partitions and Young diagrams,
\begin{align}
\lambda^{(\mathbf{i},\alpha)} \iff \{ (i,j) \vert 1 \leq i \leq l(\lambda^{(\mathbf{i}, \alpha)} ), 1 \leq j \leq \lambda^{(\mathbf{i},\alpha)} _i \},
\end{align}
and we refer them interchangeably. For each element of $\boldsymbol{\lambda} \in \mathcal{M}_\gamma (\mathbf{n}, \mathbf{k})^{T_H}$ we associate the character
\begin{align}
\mathcal{T}[\boldsymbol{\lambda}] &= \sum_{\mathbf{i} \in \text{Vert}_\gamma} \left( N_\mathbf{i} K_\mathbf{i} ^*+ q_1 q_2 N_\mathbf{i} ^* K_\mathbf{i} - (1-q_1)(1-q_2) K_\mathbf{i} K_\mathbf{i} ^* -M_\mathbf{i} ^* K_\mathbf{i} \right) \nonumber \\
&  - \sum_{\mathbf{e} \in \text{Edge}_\gamma} e^{\beta m_\mathbf{e}} (N_{t(\mathbf{e})} K_{s(\mathbf{e})} ^* +q_1 q_2 N_{s(\mathbf{e})} ^*  K_{t(\mathbf{e})} - (1-q_1)(1-q_2) K_{t(\mathbf{e})} K_{s(\mathbf{e})} ^*)  \label{charac}
\end{align}
where $q_i = e^{\beta \varepsilon_i}$ for $i=1,2$ and
\begin{align}
N_\mathbf{i} = \sum_{\alpha=1} ^{\mathpzc{n}_\mathbf{i}} e^{\beta a_{\mathbf{i},\alpha}}, \quad K_\mathbf{i} [\boldsymbol{\lambda}] = \sum_{\alpha=1} ^{\mathpzc{n}_\mathbf{i}} \sum_{\Box \in \lambda^{(\mathbf{i}, \alpha)}} e^{\beta c_{\hat{\Box}}}, \quad M_\mathbf{i} = \sum_{f=1} ^{\mathpzc{m}_\mathbf{i}} e^{\beta m_{\mathbf{i},f}} .
\end{align}
Here we are abusing the notation so that the vector spaces and the their characters under the $T_H$-action are denoted by the same letters. Also we have defined the content of the box
\begin{align} 
c_{\hat{\Box}} = a_{\mathbf{i}, \alpha} + \varepsilon_1 (i-1) + \varepsilon_2 (j-1) \quad \text{for} \quad \Box = (i,j) \in \lambda^{(\mathbf{i},\alpha)} \iff 1 \leq j \leq \lambda_i ^{(\mathbf{i},\alpha)}.
\end{align}
Finally the instanton partition function of the quiver gauge theory for $\gamma$ is 
\begin{align}
\EuScript{Z}^{\text{inst}} _\gamma ( \mathbf{a}; \textbf{\textit{m}}; \boldsymbol{\varepsilon} ; \mathfrak{q}) = \sum_{\boldsymbol{\lambda}} \prod_{\mathbf{i} \in \text{Vert}_\gamma} \mathfrak{q}_\mathbf{i} ^{\vert \boldsymbol{\lambda}^{(\mathbf{i})} \vert} \epsilon (-\mathcal{T}[\boldsymbol{\lambda}]), \label{instpart}
\end{align}
where the $\epsilon$-symbol\footnote{Not to be confused with $\varepsilon_{1,2}$ which are $\Omega$-deformation parameters.} converts a character into a product of weights
\begin{align}
\epsilon(R) = \frac{\prod_{w\in W(R^+)} w(\theta)}{\prod_{w \in W(R^-) } w(\theta)} \quad \text{for} \quad \theta \in \text{Lie}(T_H), \quad R = \sum_{w \in R^+} e^{w(\theta)} - \sum_{w \in R^-} e^{w(\theta)}.
\end{align}
Therefore we can view the instanton partition function as the partition function of the grand canonical ensemble on the colored partitions $\{ \boldsymbol{\lambda} \}$, with the measure $\boldsymbol{\mu}_{\boldsymbol{\lambda}} (\mathbf{a} , \textbf{\textit{m}}, \boldsymbol{\varepsilon}) = \prod_{\mathbf{i} \in \text{Vert}_\gamma} \mathfrak{q}_\mathbf{i} ^{\vert \boldsymbol{\lambda}^{(\mathbf{i})} \vert} \epsilon ( - \mathcal{T}[\boldsymbol{\lambda}] )$.

\subsection{$qq$-characters}
Let us first introduce the $\EuScript{Y}$-observable, which is a local observable defined as the regularized characteristic polynomial of the adjoint scalar field evaluated at $0 \in \mathbb{C}^2$ \cite{nekpes,nekpessha},
\begin{align}
\EuScript{Y}_\mathbf{i} (x) =  x^N \text{exp} \sum_{l=1} ^\infty \left[ - \frac{1}{l x^l}  \text{Tr} \phi_\mathbf{i}  ^l \vert_0  \right], \quad \mathbf{i} \in \text{Vert}_\gamma. \label{yobs}
\end{align}
As we have seen in section \ref{instpartfunc}, the supersymmetric partition function localizes on $\mathcal{M}_\gamma (\mathbf{n})^{T_H} = \bigsqcup_\mathbf{k} \mathcal{M}_\gamma(\mathbf{n}, \mathbf{k})^{T_H}$ which is the set of all colored partitions. Therefore the $\EuScript{Y}$-observable is also reduced to an observable in this statistical model on $\mathcal{M}_\gamma(\mathbf{n})^{T_H}$, expressed as
\begin{align}
\EuScript{Y}_\mathbf{i} (x)[\boldsymbol{\lambda}] = \prod_{\alpha=1} ^{\mathpzc{n}_\mathbf{i}} \left( (x-a_{\mathbf{i},\alpha}) \prod_{\Box \in \lambda^{(\mathbf{i},\alpha)}} \frac{(x- c_{\hat{\Box}} -\varepsilon_1) (x- c_{\hat{\Box}} -\varepsilon_2)}{(x- c_{\hat{\Box}} )(x- c_{\hat{\Box}} -\varepsilon)} \right), \label{yobsst}
\end{align}
where we used the notation $\varepsilon = \varepsilon_1 + \varepsilon_2$ for brevity. As observed in \eqref{yobs}, the $\EuScript{Y}$-observable is the generating functional for the gauge-invariant chiral observables
\begin{align}
\EuScript{O} _{\mathbf{i},k} = \text{Tr} \phi_\mathbf{i} ^k \vert_0. \label{chiobs}
\end{align}
Therefore the statistical model expression for the gauge-invariant chiral observables can be extracted from \eqref{yobsst} as
\begin{align}
\EuScript{O}_{\mathbf{i},k} & [\boldsymbol{\lambda}] = \sum_{\alpha=1} ^{\mathpzc{n}_\mathbf{i}} \left[ a_{\mathbf{i},\alpha} ^k + \sum_{\Box \in \lambda^{(\mathbf{i},\alpha)}} \left( (c_{\hat{\Box}} + \varepsilon_1)^k + ( c_{\hat{\Box}} + \varepsilon_2)^k - c_{\hat{\Box}} ^k - (c_{\hat{\Box}} + \varepsilon)^k \right) \right]. \label{obsst}
\end{align}
The $qq$-character is an observable defined as a certain explicitly computable Laurent polynomial $\EuScript{X}(\EuScript{Y}(x+\cdots))$ in $\EuScript{Y}$ with possibly shifted arguments \cite{nek2}. In \cite{nek4}, it was shown that the $qq$-characters can be obtained as the crossed instanton partition functions, which are the partition functions of the low-energy effective theories on intersecting branes. For example, the fundamental $qq$-character of the pure $\EuScript{N}=2$ $U(N)$ gauge theory is defined as,
\begin{align}
\EuScript{X}(\EuScript{Y})(x) = \EuScript{Y} (x+ \varepsilon) + \frac{\Lambda^{2N}}{\EuScript{Y}(x)}.
\end{align}
From the compactness theorem proven in \cite{nek3}, it has been shown that the expectation value in the gauge theory,
\begin{align}
\langle \EuScript{X}(\EuScript{Y})(x) \rangle = \frac{1}{\EuScript{Z}^{\text{inst}}} \sum_{\boldsymbol{\lambda}} \EuScript{X} (\EuScript{Y}[\boldsymbol{\lambda}]) (x) \mathfrak{q}^{\vert \boldsymbol{\lambda} \vert} \boldsymbol{\mu}_{\boldsymbol{\lambda}} (\mathbf{a},\boldsymbol{\varepsilon}) = T(x),
\end{align}
is a polynomial in $x$. Namely, the $qq$-character provides the vanishing equations for the coefficients of the negative powers of $x$: the non-perturbative Dyson-Schwinger equations.

\subsection{Bethe/gauge correspondence} \label{bethegauge}
It has been known that the low-energy effective theory of (un-deformed) four-dimensional $\EuScript{N}=2$ supersymmetric gauge theories can be described by classical integrable systems \cite{swint1, swint2}. A well-established example is the correspondence between the class-$\EuScript{S}$ theories and the Hitchin integrable systems \cite{gai1, hit1,nekpes}. Setting the 6-dimensional $\EuScript{N}=(0,2)$ superconformal theory on $\mathbb{R}^3 \times S^1 \times \EuScript{C}_{g,n}$, where $\EuScript{C}_{g,n}$ is the Riemann surface with $g$ genus and $n$ punctures, and reducing on $S^1 \times \EuScript{C}_{g,n}$ in two different orders, we observe that the total space of the fibration of the Jacobian of the Seiberg-Witten curve on the Coulomb moduli space of the class-$\EuScript{S}$ theory is identical to the phase space of the Hitchin integrable system on $\EuScript{C}_{g,n}$. The correspondence can be extended to more general four-dimensional $\EuScript{N}=2$ gauge theories with less hypermultiplets by taking proper decoupling limits. In this paper we are mainly interested in the pure $U(N)$ gauge theory. It is well-known that the corresponding integrable system is the $N$-particle periodic Toda system \cite{swint1,swint3}.

The $N$-periodic Toda system is the algebraic integrable system of $N$ non-relativistic particles in one dimension with the interaction
\begin{align}
V(x_1, \cdots, x_N) = \Lambda^2 \sum_{i=1} ^N e^{x_i - x_{i+1}},
\end{align}
and the periodicity $x_{N+1}=x_1$. The Lax operator for this system can be written as
\begin{align}
L(z) = \begin{pmatrix} p_1 & \Lambda^2 e^{x_1 - x_2} & 0 & \cdots & \cdots &  \Lambda^N z^{-1} \\ 1 & p_2 & \Lambda^2 e^{x_2 - x_3} & 0 & \cdots & 0 \\ 0 & 1 & p_3 & \Lambda^2 e^{x_3 - x_4} & \cdots  & 0 \\ 0 & \cdots & \cdots & \cdots & \cdots & 0 \\ 0 & \cdots & \cdots & \cdots & p_{N-1}& \Lambda^{2} e^{x_{N-1} - x_N} \\ \Lambda^{2-N} e^{x_N - x_1} z & 0 & \cdots & 0 & 1 & p_N \end{pmatrix},
\end{align}
from which we define the spectral curve
\begin{align}
\Sigma (x,z): \quad 0 = \text{Det} ( x- L(z)) = - \Lambda^{N} (z + z^{-1}) + x^N + u_1 x^{N-1} +u_2  x^{N-2} + \cdots + u_N. \label{speccurve}
\end{align}
The standard Lax formalism tells that the (classical) Hamiltonians,
\begin{align}
u_1 = - \sum_{i=1} ^N p_i, \quad u_2 = - \sum_{i<j} p_i p_j + \Lambda^2 \sum_i e^{x_i - x_{i+1}}, \cdots,
\end{align}
mutually commute with respect to the Poisson bracket $\{p_i, x_j \} = \delta_{ij}$, and thus establishes the classical integrability. Note that the spectral curve \eqref{speccurve} is precisely the Seiberg-Witten curve of the pure $U(N)$ gauge theory, in which $\{u_k = \langle \EuScript{O}_k \rangle \vert k = 1, \cdots, N\}$ spans the Coulomb branch of the vacua. Therefore we observe the correspondence between the low-energy description of the pure $U(N)$ gauge theory and the classical $N$-particle periodic Toda system. (See also \cite{bra,braeti} for the earlier work in the case of Toda/pure $\EuScript{N}=2$.)

In \cite{neksha3} the correspondence between the vacua of $\EuScript{N}=2$ theories and integrable systems was promoted further to the quantum level. Let us turn on the $\Omega$-deformation and take the Nekrasov-Shatashvili limit $(\varepsilon_1 \neq 0, \varepsilon_2 \rightarrow 0)$. Since we have used one of the two orthogonal rotations to deform the theory, the theory can be now effectively described as a two-dimensional theory with $\EuScript{N}=(2,2)$ supersymmetry. The low-energy effective action of this two-dimensional theory contains the twisted $F$-term\footnote{\label{foot1}In the reduction from the four-dimensional $\EuScript{N}=2$ to the two-dimensional $\EuScript{N}=(2,2)$, we are choosing the convention in which $\EuScript{N}=(2,2)$ gauge multiplet is described by the twisted chiral superfield. Note that the complex adjoint scalar in the $\EuScript{N}=2$ vector multiplet becomes the one in the $\EuScript{N}=(2,2)$ twisted chiral multiplet under this reduction. See section \ref{partial}.} from the effective twisted superpotential $\widetilde{\EuScript{W}} (\mathbf{a} , \varepsilon_1, \mathfrak{q})$, which can be computed by the supersymmetric localization for generic $(\mathbf{a}, \varepsilon_1)$ as
\begin{align}
\EuScript{\widetilde{W}} (\mathbf{a} , \varepsilon_1, \mathfrak{q}) = \lim_{\varepsilon_2 \to 0} \varepsilon_2 \text{log} \EuScript{Z} (\mathbf{a} , \boldsymbol{\varepsilon}, \mathfrak{q}). \label{efftwsup}
\end{align}
The effective twisted superpotential becomes important for determining vacua and expectation values of the twisted chral observables, as we shall see below.

The space of vacua of the effective theory is a representation of the twisted chiral ring, which is spanned by the gauge-invariant polynomials of the complex adjoint scalar, \eqref{chiobs}.\footnote{See footnote \ref{foot1} and section \ref{partial}. The \textit{chiral} observables in the four-dimensional gauge theory are reduced to the \textit{twisted chiral} observables in the effective two-dimensional theory.} In \cite{neksha1,neksha2}, it was shown that the twisted chiral ring of a two-dimensional $\EuScript{N}=(2,2)$ gauge theory is identified with the Hamiltonians of the corresponding integrable system. Namely, the problem of quantization becomes the spectral problem, with the identification
\begin{align}
\langle \EuScript{O}_k \rangle \vert_{\varepsilon_2 \rightarrow 0 , \mathbf{a} \in vac} = E_k(\mathbf{a},\varepsilon_1),
\end{align}
the eigenvalue of the corresponding quantum Hamiltonian $\hat{H}_k$. Here the equation for the vacua of the two-dimensional effective theory corresponds to the quantization condition of the integrable system. As noted in \cite{neksha3, nekwit}, the Nekrasov-Shatashvili limit of the $\EuScript{N}=2$ supersymmetric gauge theory leads to several quantization conditions and correspondingly to different quantum integrable systems. The choice of quantization condition becomes manifest in the topological sigma model description of the quantization. We can interpret the Nekrasov-Shatashvili limit of the $\Omega$-deformation as the cigar metric $\mathbb{R} \times S^1 \times D_R$, in which the cigar has the asymptotic behavior of $D_R \sim I \times S^1$ with $I = [0,R]$. Then by reducing the four-dimensional $\EuScript{N}=2$ gauge theory on $\mathbb{R} \times I$, the theory is reduced to the topological A-model with the worldsheet with the boundaries and the target space being the complexified phase space. We can make use of the brane quantization picture from this topological A-model description \cite{gukwit}. In particular, the quantization is realized by choosing the boundary condition at $0 \in I$ to be the canonical coisotropic $A$-brane and the boundary condition at $R \in I$ to be the Lagrangian $A$-brane. There are two classes of the Lagrangian $A$-branes that can be chosen, which lead to two different types of the quantization:
\begin{subequations}
\begin{align}
\text{Type A:}& \quad \quad \text{exp} \left( 2\pi \frac{\partial \EuScript{\widetilde{W}}}{\partial a_\alpha} - i \theta_\alpha \right)=1, \label{typeA} \\
\text{Type B:}& \quad \quad \text{exp} \left( 2 \pi i \frac{a_\alpha}{\varepsilon_1} - i \theta_\alpha \right) =1, \quad \theta_\alpha \in [0, 2\pi). \label{typeB}
\end{align}
\end{subequations}
In the original four-dimensional gauge theory on $\mathbb{R} \times S^1 \times D_R$, they correspond to the choices of the supersymmetric boundary conditions at $R \in I$. In particular, the type A condition corresponds to the Neumann boundary condition for the vector multiplet. In this case the four-dimensional vector multiplet is reduced to the two-dimensional vector multiplet in the effective theory on $\mathbb{R} \times S^1$, which is $\EuScript{N}=(2,2)$ abelian gauge theory so that the vacua are determined by the effective twisted superpotential as \eqref{typeA} (we included the $\theta$-shift). The type B condition corresponds to the Dirichlet condition for the vector multiplet. The gauge symmetry is completely broken and both vector multiplets and hypermultiplets of the four-dimensional theory are reduced to chiral multiplets of the effective two-dimensional theory. We impose the vanishing condition for the holonomy around the boundary $\partial D_R$ to preserve the supersymmetry, yielding the quantization condition \eqref{typeB}. See \cite{nekwit} for more detail.

For the case of the pure $U(N)$ gauge theoy, type A and B reality conditions correspond to the following formulations of quantum periodic Toda system. In the type A quantization, we are taking the real slice of $x_i \in \mathbb{R}$. After decoupling the motion of the center of mass, we look for the $L^2$-normalizable eigenfunctions with real and discrete spectra. It was shown that the vacuum equation \eqref{typeA} precisely leads to the Gutzwiller quantization condition for this type of spectral problem \cite{koztes}. See also \cite{gut,gp,kl1,kl2,an} for previous works on the type A periodic Toda system.

In this paper, we mainly focus on the type B quantization of the periodic Toda system, which shows quite a different interesting feature. Here we have (quasi-)periodic eigenfunctions with the period $2\pi i$. The spectra of the Hamiltonians are complex but still discrete. With the $\theta$-shift, the quantization condition is
\begin{align}
a_\alpha = \left( n_\alpha + \frac{\theta_\alpha}{2 \pi} \right) \varepsilon_1, \quad n_\alpha \in \mathbb{Z}, \label{quantcon}
\end{align}
where $\theta_\alpha$ is precisely the Bloch angle for the shift of $x_\alpha$ by the period $2 \pi i$. The spectra of the Hamiltonians can be computed as the expectation value of the observables in the twisted chiral ring, under the Nekrasov-Shatashvili limit with the condition \eqref{quantcon} imposed:
\begin{align}
E_k (\mathbf{a}, \varepsilon_1) = \langle \EuScript{O}_k \rangle \vert_{\varepsilon_2 \rightarrow 0, \eqref{quantcon}} = \frac{1}{\EuScript{Z}^{\text{inst}}} \sum_{\boldsymbol{\lambda}} \mathfrak{q}^{\vert \boldsymbol{\lambda} \vert} \EuScript{O}_k [\boldsymbol{\lambda}] \boldsymbol{\mu}_{\boldsymbol{\lambda}}(\mathbf{a}, \boldsymbol{\varepsilon}) \Bigg \vert_{\varepsilon_2 \rightarrow 0, \eqref{quantcon}},
\end{align}
where the statistical model form of the observable $\EuScript{O}_k [\boldsymbol{\lambda}]$ is given in \eqref{obsst}. In particular, the spectra of two lowest order Hamiltonians $\EuScript{O}_2$ and $\EuScript{O}_3$ take simple form:
\begin{subequations} \label{energies}
\begin{align}
E_2 (\mathbf{a},\varepsilon_1) &= \left[ \sum_\alpha a_\alpha^2 - \frac{1}{N} \varepsilon_1 \Lambda \frac{\partial \widetilde{\EuScript{W}}}{\partial \Lambda} \right]\Bigg\vert_\eqref{quantcon}, \label{energy} \\
E_3 (\mathbf{a},\varepsilon_1) &= \left[ \sum_\alpha a_\alpha ^3 - \frac{3\varepsilon_1 ^2}{2N} \Lambda \frac{\partial \widetilde{\EuScript{W}}}{\partial \Lambda} - 6 \varepsilon_1 \lim_{\varepsilon_2 \to 0} \varepsilon_2 \Bigg\langle \sum_{\Box \in K} c_{\hat{\Box}} \Bigg\rangle \right]\Bigg\vert_\eqref{quantcon}. \label{energy3}
\end{align}
\end{subequations}
For example, in the case of $N=2$ the type B quantum periodic Toda system is reduced to the Mathieu system, whose discrete energy spectrum has been well-studied. For generic value of the Coulomb moduli $a_{12}=a_1 - a_2$ (on the integrable system side, generic value of $\theta_1 - \theta_2$) the gauge theory computation of the spectrum \eqref{energy} precisely reproduces the known perturbative computation, order by order in the series of $\Lambda^4$. We also checked that the perturbative spectra of $N=3$ periodic Toda system are reproduced by \eqref{energies}. The generalization of the computation to the higher $N$ is straightforward.

However, when the Coulomb moduli assume special values $\frac{a_{\alpha \beta}}{\varepsilon_1} \in \mathbb{Z} \setminus \{0\}$, the correspondence breaks down as we now describe. A relation among the equivariant parameters implies that the maximal torus $T_H$ used for the equivariant localization becomes smaller than generic cases. When the torus becomes smaller, the set of fixed points $\EuScript{M} (N) ^{T_H}$ in general becomes larger; as noted in \cite{nek2}, one may find a copy of $\mathbb{P}^1$'s or a even more complicated subvariety instead of isolated set of fixed points with the reduction of symmetry group. 

It can be shown that for the specific case at hand, $\frac{a_{\alpha \beta}}{\varepsilon_1} \in \mathbb{Z} \setminus \{0\}$, $\EuScript{M} (N) ^{T_H}$ actually contains products of $\mathbb{P}^1$'s. Recall that before taking $\frac{a_{\alpha \beta}}{\varepsilon_1} \in \mathbb{Z} \setminus \{0\}$ the isolated fixed points $\EuScript{M} (N) ^{T_H}$ are classified by $N$-tuples of Young diagrams $\{\boldsymbol\lambda\}$. The boxes in these Young diagrams encode the weights of linearly independent vectors in the space $K[\boldsymbol\lambda]$ in terms of Coulomb moduli and $\Omega$-deformation parameters, and these weights are all distinct. However, once we introduce the new constraint $\frac{a_{\alpha \beta}}{\varepsilon_1} \in \mathbb{Z} \setminus \{0\}$, the weights now may overlap (or in terms of the Young diagrams, two boxes in different Young diagrams may collide). This implies two isolated fixed points disappear into an emergent fixed point set $\mathbb{P}^1$ (so that when the symmetry group action is refined by an extra $U(1)$ as it used to be, we recover two isolated fixed points on the emergent $\mathbb{P}^1$). Since we get an emergent $\mathbb{P}^1$ whenever this overlap occurs, the fixed point set $\EuScript{M} (N) ^{T_H}$ now contains a product of mutiple $\mathbb{P}^1$'s.

Hence the integral that provides the instanton partition function remains finite due to the compactness of $\EuScript{M} (N) ^{T_H}$. Nevertheless, the integral over the emergent $\mathbb{P}^1$'s gives additional poles in $\varepsilon_2$, altering the asymptotic behavior of the instanton partition function in the limit $\varepsilon_2 \to 0$. Most importantly, the effective twisted superpotential is not properly obtained by taking $\widetilde{\EuScript{W}} = \lim_{\varepsilon_2 \to 0} \varepsilon_2 \text{log} \EuScript{Z}$ since the expression becomes divergent. Therefore we see that \eqref{energies} cannot work as it is stated. The main subject of the present work is to recover the correspondence at this special locus.

\section{Gauge theory with partial $\Omega$-deformation and partial noncommutativity} \label{partial}
To explore the gauge theoretical meaning of the special locus of Coulomb moduli, let us study the pure $\EuScript{N}=2$ $U(N)$ gauge theory with partial $\Omega$-deformation and partial noncommutativity. The four-dimensional $\EuScript{N}=2$ supersymmetry can be described by the super-covariant derivatives in the covariant basis,
\begin{align} 
&\{ {\boldsymbol{\nabla}}_\alpha ^{A} , {\bar{\boldsymbol{\nabla}}_{B \dot{\alpha}}} \} = - i {\delta^A}_B \boldsymbol{\nabla}_{\alpha \dot{\alpha}} \nonumber \\
&\{ {\boldsymbol{\nabla}}_\alpha ^{A} , {\boldsymbol{\nabla}}_\beta ^{B} \} = i \epsilon^{AB} \epsilon_{\alpha \beta} \bar{\boldsymbol{\Phi}} \nonumber \\
&\{ \bar{{\boldsymbol{\nabla}}}_{A \dot{\alpha}} , \bar{{\boldsymbol{\nabla}}} _{B \dot{\beta}} \} = i \epsilon_{AB} \epsilon_{\dot{\alpha} \dot{\beta}} {\boldsymbol{\Phi}},
\end{align}
where we are using the convention $\sigma_{ \alpha \dot{\alpha}} ^\mu = ( \mathds{1}_{\alpha \dot{\alpha} },\vec{\tau} _{  \alpha \dot{\alpha} })$. Here $\boldsymbol{\Phi}$ is the $\EuScript{N}=2$ chiral superfield constrained by $\boldsymbol{\nabla}^\alpha _A \boldsymbol{\nabla}_{B \alpha} \boldsymbol{\Phi} = - \bar{\boldsymbol{\nabla} } _{B\dot{\alpha}} \bar{\boldsymbol{\nabla}}_{A } ^{\dot{\alpha}} \bar{\boldsymbol{\Phi}} $ due to the Bianchi identities. The action for the pure $\EuScript{N}=2$ gauge theory can be written in the $\EuScript{N}=2$ chiral superspace as
\begin{align}
\mathcal{L} = \frac{1}{8\pi} \text{Im} \int  d^4 \theta \frac{1}{2} \tau \text{Tr} \boldsymbol{\Phi} ^2. \label{n2supac}
\end{align}
The partial $\Omega$-deformation ($\varepsilon_1 \neq 0$, $\varepsilon_2=0$) breaks the $\EuScript{N}=2$ supersymmetry, but preserves a $\EuScript{N}=(2,2)$ subalgebra on the $(x^0, x^3)$-plane,
\begin{align}
\{ {\boldsymbol{\nabla}}_+ ^{1} , {\bar{\boldsymbol{\nabla}}_{1 \dot{+}}} \} &= - i \boldsymbol{\nabla}_{+ \dot{+} } = -i \left( \boldsymbol{\nabla}_{0} + \boldsymbol{\nabla}_{3} \right) \nonumber \\
\{ {\boldsymbol{\nabla}}_- ^{2} , {\bar{\boldsymbol{\nabla}}_{2 \dot{-}}} \} &= - i \boldsymbol{\nabla}_{-\dot{-} } = - i \left( \boldsymbol{\nabla}_{0} - \boldsymbol{\nabla}_{3} \right).
\end{align}
Let us choose the following convention for the reduced algebra
\begin{align}
&\boldsymbol{\nabla}_+ ^{1} \equiv \boldsymbol{\nabla}_+, \quad \boldsymbol{\nabla}_- ^{2} \equiv  \bar{\boldsymbol{\nabla}}_-, \nonumber \\
&\bar{\boldsymbol{\nabla}}_{1 \dot{+}}  \equiv  \bar{\boldsymbol{\nabla}}_+, \quad \bar{\boldsymbol{\nabla}}_{2 \dot{-}} \equiv \boldsymbol{\nabla}_-,
\end{align}
so that the restriction of the $\EuScript{N}=2$ chiral superfield $\Sigma \equiv   \boldsymbol{\Phi} \vert = i  \{ \bar{\boldsymbol{\nabla}}_+ , \boldsymbol{\nabla}_- \}$ is a twisted chiral superfield in the reduced $\EuScript{N}=(2,2)$ supersymmetry. Note that $\Sigma$ contains the complex scalar of the $\EuScript{N}=2$ vector multiplet as its component field. Also it is important that we have the following relations from the Bianchi identities,
\begin{align}
[\bar{\boldsymbol{\nabla}}_\pm , \boldsymbol{\nabla}_{- \dot{+}}] =0. \label{rel}
\end{align}
The $\EuScript{N}=2$ superspace action is reduced to the $\EuScript{N}=(2,2)$ superspace,
\begin{align}
\mathcal{L} = -\frac{1}{2 g^2} \int d^4 \theta \text{Tr} \bar{\Sigma}\Sigma - \text{Im} \left[ \frac{\tau}{8\pi} \int d^2 \tilde{\theta}  \text{Tr}\left(  i\Sigma [\boldsymbol{\nabla}_{+ \dot{-}} , \boldsymbol{\nabla}_{- \dot{+}} ] - [\boldsymbol{\nabla}_- , \boldsymbol{\nabla}_{- \dot{+}}][\bar{\boldsymbol{\nabla}}_+ , \boldsymbol{\nabla}_{+ \dot{-}}] \right) \right].
\end{align}
Now let us turn on the noncommutativity on the $(x^1, x^2)$-plane, $[x^1 , x^2] = i \zeta$, while leaving the $(x^0, x^3)$-plane commutative. Define the raising and the lowering operators:
\begin{align}
c = \frac{1}{\sqrt{2 \zeta}}& (x^1 + i x^2), \quad c^\dagger = \frac{1}{\sqrt{2 \zeta}} (x^1 - i x^2), \quad [c,c^\dagger] =1. \label{noncomalg}
\end{align}
The effect of the noncommutativity is that the \textit{covariant coordinate}
\begin{align}
\Phi \equiv  - i \frac{1}{\sqrt{\zeta}} c  - \frac{1}{\sqrt{2}} ( A_1 +  i A_2)
\end{align}
can act by commutator as the covariant derivative along the noncommutative direction \cite{Cor}. Namely, we can make a substitution $\boldsymbol{\nabla}_{- \dot{+}} \to \sqrt{2} \Phi $ except in the commutator of two such covariant derivatives,
\begin{align}
[\boldsymbol{\nabla}_{- \dot{+}} , \boldsymbol{\nabla} _{+ \dot{-}} ] = 2 [\Phi, \bar{\Phi}]- \frac{2}{\zeta},
\end{align}
where we have the extra term from the commutator of $c$ and $c^\dagger$. Note that $\Phi$ is an adjoint chiral superfield in the $\EuScript{N}=(2,2)$ supersymmetry by the relation \eqref{rel}. The fields are now promoted to endomorphisms of the Fock space $\mathcal{H}$ that represents the algebra \eqref{noncomalg}, on which the dependence of the fields on the noncommutative coordinates are encoded. The integration along the noncommutative directions is replaced by the trace over the Fock space,
\begin{align}
\int dx^1 dx^2 (\cdots) = \zeta \text{Tr}_\mathcal{H} (\cdots).
\end{align}
Thus, with the Wick rotation, we arrive at the Euclidean two-dimensional $\EuScript{N}=(2,2)$ superspace action of the four-dimensional theory with the partial noncommutativity
\begin{align}
\mathcal{L} = \frac{i}{ 8\pi} \left(  \tau \int d^2 \tilde{\theta} \text{Tr}_{\mathcal{H} \otimes \mathbb{C}^N} \Sigma + c.c \right) + \frac{\zeta}{g^2} \int d^4 \theta \text{Tr}_{\mathcal{H} \otimes \mathbb{C}^N} \left[ - \frac{1}{2} \bar{\Sigma} \Sigma +  \bar{\Phi} e^{V } \Phi \right],
\end{align}
with the following superfield contents\footnote{Here we are denoting the complex scalar which descends from the $\EuScript{N}=2$ vector multiplet as $\sigma$, which has been denoted as $\phi$ so far. The convention may be confusing but is more traditional in $\EuScript{N}=(2,2)$ context.}
\begin{subequations}
\begin{align}
\text{Twisted chiral :}&\quad \Sigma = (\sigma, \lambda_+, \bar{\lambda}_-, iD+ F_{43}) \\
\text{Adjoint chiral :}&\quad \Phi = (\phi, \psi_{\pm} , F).
\end{align}
\end{subequations}
As is apparent from the definition of $\Phi$ as the covariant coordinate, the $U(1) = SO(2)_{12} \subset G_\text{rot}$ spacetime rotation becomes the flavor symmetry rotating the chiral multiplet $\Phi$. The partial $\Omega$-deformation $(\varepsilon_1 \neq 0, \varepsilon_2 =0)$ is simply weakly gauging this $U(1)$ flavor symmetry to generate the twisted mass for the chiral multiplet,
\begin{align}
\widetilde{V}_{\varepsilon_1} = -\varepsilon_1 \theta^- \bar{\theta}^+ - \bar{\varepsilon}_1 \theta^+ \bar{\theta}^-.
\end{align}
Thus the final form of the action is 
\begin{align}
\mathcal{L} = \frac{i}{8\pi} \left(  \tau \int d^2 \tilde{\theta} \text{Tr}_{\mathcal{H} \otimes \mathbb{C}^N} \Sigma + c.c \right) + \frac{\zeta}{g^2} \int d^4 \theta \text{Tr}_{\mathcal{H} \otimes \mathbb{C}^N} \left[ -\frac{1}{2}  \bar{\Sigma} \Sigma +  \bar{\Phi} e^{V + \widetilde{V}_{\varepsilon_1}} \Phi \right], \label{superspace}
\end{align}
which can be expanded to an $x$-space action,
\begin{align}
\mathcal{L} = \frac{\zeta}{g^2} \text{Tr}_{\mathcal{H} \otimes \mathbb{C}^N} & \left[ \frac{1}{2} F_{43}^2 + D_\mu \sigma^\dagger D^\mu \sigma + \frac{1}{2} D^2 -i D([\phi, \phi^\dagger] -  \frac{1}{\zeta}) + \frac{1}{2} [\sigma , \sigma^\dagger]^2 + FF^\dagger \right.  \nonumber \\
& + D_\mu \phi^\dagger D^\mu \phi + \vert [\sigma , \phi] + \varepsilon_1 \phi \vert ^2 +  \vert [\sigma , \phi^\dagger] - \varepsilon_1 \phi^\dagger \vert ^2 \nonumber \\
& +2i \bar{\lambda}_+ D_{z} \lambda_+ - 2 i \bar{\lambda}_- D_{\bar{z}} \lambda_- + 2i \bar{\psi}_+ D_{z} \psi_+ - 2 i \bar{\psi}_- D_{\bar{z}} \psi_- \nonumber \\
& + \sqrt{2} \lambda_+ [\sigma, \bar{\lambda}_-] - \sqrt{2} [\sigma^\dagger , \lambda_- ] \bar{\lambda}_+ + \sqrt{2} \bar{\psi}_+ ([\sigma^\dagger , \psi_-] + \bar{\varepsilon}_1 \psi_-) + \sqrt{2} \bar{\psi}_- ( [\sigma , \psi_+ ] + \varepsilon_1 \psi_+) \nonumber \\ 
& \left. - i\sqrt{2} \bar{\psi}_+ [\bar{\lambda}_- , \phi] + i \sqrt{2} \bar{\psi}_- [\bar{\lambda}_+ , \phi] - i \sqrt{2} [\phi^\dagger, \lambda_+] \psi_- + i \sqrt{2} [\phi^\dagger , \lambda_-] \psi_+ \right] \nonumber \\
- \frac{i\vartheta}{8 \pi^2}& \text{Tr}_{\mathcal{H} \otimes \mathbb{C}^N} F_{43}. \label{action}
\end{align}
The bosonic part of the action can be written as
\begin{align}
\mathcal{L}_\text{bos}& = - \frac{i \tau}{4\pi}  \text{Tr}_{\mathcal{H} \otimes \mathbb{C}^N} F_{43} \nonumber \\
& \quad + \frac{\zeta}{g^2} \text{Tr}_{\mathcal{H} \otimes \mathbb{C}^N} \left[ \frac{1}{2} \left(F_{43} + [\phi, \phi^\dagger] - \frac{1}{\zeta} \right)^2 + 4\vert D_{\bar{z}} \phi \vert^2 + D_\mu \sigma^\dagger D^\mu \sigma +F F^\dagger  \right. \nonumber \\
& \quad \quad \left. + \frac{1}{2} \left( D- i \left( [\phi,\phi^\dagger] - \frac{1}{\zeta} \right) \right)^2 +  \vert [\sigma, \phi] + \varepsilon_1 \phi \vert^2 + \vert [\sigma, \phi^\dagger] - \varepsilon_1 \phi^\dagger \vert^2 + \frac{1}{2} [\sigma, \sigma^\dagger]^2  \right],
\end{align}
from which we read off the vaccum equations
\begin{align}
F_{43} &+ [\phi , \phi^\dagger] - \frac{1}{\zeta} =0, \quad D_{\bar{z}} \phi =0, \quad D- i \left([\phi, \phi^\dagger] - \frac{1}{\zeta} \right) =0,\nonumber \\
D_\mu& \sigma =0, \quad [\sigma, \sigma^\dagger] = 0, \quad [\sigma , \phi ] + \varepsilon_1 \phi = [\sigma, \phi^\dagger] - \varepsilon_1 \phi^\dagger =0. \label{vac}
\end{align}
We focus on the trivial sector where $F_{43}=0$. Then the vaccum equations are solved by
\begin{align}
D&=0, \nonumber \\
\sigma &= \varepsilon_1 c^\dagger c \otimes \mathds{1}_{\mathbb{C}^N} + \mathds{1}_{\mathcal{H}} \otimes \text{diag}(a_1, a_2, \cdots, a_N), \nonumber \\
\phi &= \frac{1}{\sqrt{\zeta}} c \otimes \mathds{1}_{\mathbb{C}^N}, \quad \phi^\dagger =\frac{1}{\sqrt{\zeta}} c^\dagger \otimes \mathds{1}_{\mathbb{C}^N}, \label{vev}
\end{align}
where $a_\alpha$ are moduli that parametrize the vacua. Since $\sigma$ is the complex scalar in the $\EuScript{N}=2$ vector multiplet, $a_\alpha$ are nothing but the Coulomb moduli in the four-dimensional perspective. The low-energy effective action is obtained by integrating out all the massive modes and high energy modes around the vaccum \eqref{vev}. Thus we split the vacuum expectation value and the quantum fluctuation,
\begin{align}
\sigma = \sigma_0 + \hat{\sigma}, \quad \phi = \phi_0 + \hat{\phi},
\end{align}
and expand the action in fluctuation modes. We introduce the following gauge fixing term
\begin{align}
\mathcal{L}_{\text{fix}} = \frac{\zeta}{2 g^2} \text{Tr}_{\mathcal{H} \otimes \mathbb{C}^N} \left[ \partial_\mu A^\mu -i[\sigma_0 ^\dagger, \hat{\sigma} ] -i[\sigma_0, \hat{\sigma}^\dagger ]-i[\phi_0 ^\dagger , \hat{\phi}]-i[\phi_0, \hat{\phi}^\dagger] \right]^2,
\end{align}
to cancel the mixing terms in the quadratic order. Then we are left with
\begin{align}
\mathcal{L}&_\text{bos} + \mathcal{L}_{\text{fix}} \nonumber \\
=& \frac{\zeta}{g^2} \text{Tr}_{\mathcal{H} \otimes \mathbb{C}^N} \left[ \frac{1}{2} F_{43} ^2 + \vert [A_\mu , \sigma_0] \vert^2 +  \vert [A_\mu , \phi_0] \vert^2 + D_\mu \hat{\sigma}^\dagger D^\mu \hat{\sigma}+D_\mu \hat{\phi}^\dagger D^\mu \hat{\phi} + \frac{1}{2} (\partial_\mu A^\mu)^2 +F F^\dagger \right. \nonumber \\ 
& \quad \quad \quad\quad -i D [\hat{\phi} , \hat{\phi} ^\dagger] +\frac{1}{2} \left( D- i ([\phi_0 , \hat{\phi} ^\dagger ]+[\hat{\phi} , \phi_0 ^\dagger]) \right)^2 + 2\vert [\hat{\phi} , \phi_0 ^\dagger ] \vert^2 + \frac{1}{2} [\hat{\sigma} , \hat{\sigma} ^\dagger]^2 \nonumber \\
& \quad \quad \quad\quad+ [\hat{\sigma} , \hat{\sigma} ^\dagger] \left( [\sigma_0 , \hat{\sigma} ^\dagger] + [\hat{\sigma} , \sigma_0 ^\dagger ] \right) + 2 \vert [\hat{\sigma} , \sigma_0 ^\dagger ] \vert ^2 -[A_\mu, \sigma_0 ^\dagger][A^\mu , \hat{\sigma}] - [A_\mu, \hat{\sigma} ^\dagger][A^\mu , \sigma_0] \nonumber \\
&\quad \quad \quad\quad -[A_\mu , \phi_0 ^\dagger][A^\mu , \hat{\phi}] - [A_\mu, \hat{\phi} ^\dagger][A^\mu , \phi_0] +\vert [\sigma, \hat{\phi}] + \varepsilon_1 \hat{\phi} \vert ^2 +\vert [\sigma, \hat{\phi} ^\dagger] - \varepsilon_1 \hat{\phi} ^\dagger \vert^2 \nonumber \\
&\left.\quad \quad \quad\quad +\vert [\hat{\sigma}, \phi_0] \vert^2 + \vert[ \hat{\sigma}, \phi_0 ^\dagger] \vert^2 + [\hat{\sigma}, \hat{\phi}][\phi_0 ^\dagger , \hat{\sigma} ^\dagger] + [\hat{\phi} ^\dagger, \hat{\sigma}^\dagger][\hat{\sigma}, \phi_0] + [\hat{\sigma}, \hat{\phi} ^\dagger][\phi_0, \hat{\sigma} ^\dagger] + [\hat{\phi}, \hat{\sigma} ^\dagger][\hat{\sigma} , \phi_0 ^\dagger] \right] \nonumber \\
& \quad  - \frac{i \vartheta}{8 \pi^2} \text{Tr}_{\mathcal{H} \otimes \mathbb{C}^N} F_{43}
\end{align}
For generic values of Coulomb moduli, the only massless fluctuations are the modes of the abelian twisted chiral multiplet,
\begin{align}
\hat{\Sigma} = \hat{\sigma} + \cdots = \mathds{1}_\mathcal{H} \otimes \text{diag}(\Sigma_1, \Sigma_2, \cdots, \Sigma_N).
\end{align}
All the other modes are integrated out in the effective theory, possibly contributing to the effective twisted superpotential $\widetilde{\EuScript{W}}(\Sigma_\alpha)$. Therefore the effective two-dimensional theory is a pure abelian gauge theory of rank $N$ with a certain effective twisted superpotential. 

However, we discover that additional massless modes emerge at the special locus of Coulomb moduli, $\{  a_{\alpha \beta} = m \varepsilon_1 \;\vert\; m \in \mathbb{Z} \setminus \{0\} \}$. Namely, the mass term for the chiral multiplet mode
\begin{align}
\hat{\Phi} \equiv \hat{\phi} + \cdots = \begin{cases}  (c^\dagger)^{m-1} \otimes E_{\beta, \alpha} \Phi_{\alpha \beta}, \quad \text{if} \quad m >0 \\ (c^\dagger)^{-m-1} \otimes E_{\alpha, \beta} \Phi_{\alpha \beta}, \quad \text{if}\quad m <0 \end{cases}
\end{align}
vanishes at the locus. Here, $E_{\alpha, \beta}$ is the $N \times N$ matrix whose elements are all 0 except 1 for the element in the $\alpha$th row and the $\beta$th column. A massless mode of chiral multiplet is generated for each such a pair of $(\alpha, \beta)$. The emergent massless modes signify the failure of the effective description of the theory. In \cite{gms}, it was argued that this failure is cured by the appearence of solitonic particles, which prevent the massless modes to occur through the wall-crossing. It would be nice to directly see how this wall-crossing phenomenon interplays with the insertion of surface defects discussed in the following sections.

\section{Surface defect} \label{surface}

\subsection{Construction}
As non-local gauge-invariant observables, the surface defects enrich the study of $\EuScript{N}=2$ supersymmetric gauge theories and Bethe/gauge correspondence. There are two ways of constructing the half-BPS surface defects in the context of the $\EuScript{N}=2$ gauge theory. One of them is orbifolding the four-dimesional spacetime with respect to the action of the cyclic group $\mathbb{Z}_p$ as $\mathbb{C}_{\varepsilon_1} \times (\mathbb{C}_{\varepsilon_2} / \mathbb{Z}_p)$. This type of surface defect is referred as the orbifold surface defect. The second way is inserting a degenerate gauge vertex in the quiver which defines the quiver gauge theory of interest. Even though these constructions seem to be distinct, there is an IR duality (at least in the $A_1$ case) between the two types of surface defect that descends from the M-theory brane transition \cite{frgutes}. We introduce them both constructions below, although we mainly utilize the orbifold surface defect for our purpose. More general discussions on half-BPS surface defects on quiver gauge theories are in \cite{nek7}.
\subsubsection{Orbifold construction}
Throughout the discussion, let us restrict our attention to the pure $U(N)$ gauge theory. The orbifold surface defect $\mathcal{D}_{\mathbb{Z}_p, \rho}$ is constructed by specifying the embedding
\begin{align}
\rho : \mathbb{Z}_p \longrightarrow H = G_{g} \times G_{\text{rot}},
\end{align}
from which we define the surface defect as the prescription of performing the path integral over the space of $\mathbb{Z}_p$-invariant fields. The rotation group part of the embedding is always chosen to be 
\begin{align}
\Omega(\zeta) : (z_1 , z_2 ) \mapsto (z_1 , \zeta z_2), \quad \text{for} \quad \zeta = \text{exp}\left( \frac{2\pi i}{p} \right). \label{lorentz}
\end{align}
To fully characterize the surface defect we need to further specify the gauge group part of the embedding $\rho$. It is assigned by the coloring function
\begin{align}
c: [N]= \{0,\cdots, N-1\} \longrightarrow \mathbb{Z}_p,
\end{align}
from which we define the gauge group part of the embedding $\rho$ such that the vector space $N$ decomposes as
\begin{align}
N=\sum_{\alpha} e^{\beta a_\alpha} \EuScript{R}_{c(\alpha)} = \sum_{\omega \in \mathbb{Z}_p} N_\omega \EuScript{R}_\omega \quad \Longrightarrow \quad N_\omega = \sum_{\alpha \in c^{-1} (\omega)} e^{\beta a_\alpha},
\end{align}
where $\mathcal{R}_\omega$ is the one-dimensional irreducible representation of $\mathbb{Z}_p$ of weight $\omega$, 
\begin{align} \mathbb{Z}_p &\longrightarrow \text{End}(\mathcal{R}_\omega) \nonumber \\
\zeta &\longmapsto \zeta^\omega.
\end{align} 
Then we also decompose
\begin{align}
K = \sum_{\omega \in \mathbb{Z}_p} K_\omega \EuScript{R}_\omega, \quad \text{where} \quad
K_\omega = \sum_{\alpha}  \sum_{\substack{(i,j) \in \lambda^{(\alpha)} \\ c(\alpha) + j-1 \equiv \omega \text{ mod }p}} e^{\beta(a_\alpha + \varepsilon_1 (i-1) + \varepsilon_2 (j-1))}.
\end{align} 
We can identify the spacetime $\mathbb{C}^2$ with the orbifold $\mathbb{C}^2 /\mathbb{Z}_p$ through the map $(z_1 , z_2) \mapsto (\tilde{z}_1 = z_1 , \tilde{z}_2 = z_2 ^p)$. This map is singular along the surface $z_2 = 0$. Therefore the path integral over the space of the $\mathbb{Z}_p$-invariant fields on $(z_1 , z_2)$-space is interpreted as the path integral over the $(\tilde{z_1} , \tilde{z_2})$-space with the insertion of a defect along the surface $\tilde{z_2} =0$.

An orbifold surface defect is called \textit{regular} for the special case when $p=N$ and $c \in S_N$, where $S_N$ is the permutation group of $[N]=\{0, \cdots N-1 \}$. This special kind of surface defects plays an important role in constructing the eigenstate wavefunctions of the integrable system in section \ref{a1orb} and section \ref{splitting}.

\subsubsection{$\EuScript{N}=2$ supersymmetric gauge theory with orbifold surface defect}
We now investigate the $\EuScript{N}=2$ gauge theory in the presence of the orbifold surface defect. In the presence of the surface defect, the coupling constant is fractionalized
\begin{align}
\mathfrak{q} \mapsto \mathfrak{q}_\omega \equiv \Lambda^2 \frac{z_\omega}{z_{\omega-1}}, \quad \omega \in \mathbb{Z}_p, \label{fraccoup}
\end{align}
with $z_{\omega + p} \equiv z_\omega$. The surface defect partition function is the path integral over the space of $\mathbb{Z}_p$-invariant fields, which can be easily obtained from the bulk partition function. From \eqref{instpart}, the instanton part of the surface defect partition function is immediately obtained
\begin{align}
\boldsymbol{\Psi}^{\text{inst}}_c(\mathbf{a},\boldsymbol{\varepsilon},\mathfrak{q}, \mathbf{z}) =  \sum_{\boldsymbol{\lambda}} \prod_{\omega \in \mathbb{Z}_p}  \mathfrak{q}_\omega ^{k_\omega} \epsilon(-\mathcal{T}[\boldsymbol{\lambda}] ^{\mathbb{Z}_p , c}), \label{surfdefpart}
\end{align}
where $k_\omega [\boldsymbol{\lambda}] = \text{dim} K_\omega [\boldsymbol{\lambda}]$ is the fractionalized instanton number and $(\cdots)^{\mathbb{Z}_p , c}$ is the prescription of keeping the $\mathbb{Z}_p$-invariant piece for the given coloring function $c$ only. The $\mathbb{Z}_p$-invariant piece of the character \eqref{charac} is given by
\begin{align}
\EuScript{T}[\boldsymbol{\lambda}] ^{\mathbb{Z}_p , c} = \sum_{\omega \in \mathbb{Z}_p} \left[ N_\omega K_\omega ^* +q_1 q_2 N_\omega ^* K_{\omega-1} - (1-q_1 ) K_\omega K_\omega ^* + q_2(1-q_1) K_\omega K_{\omega+1} ^* \right].
\end{align}
In the special case that the coloring function $c: [N] \to \mathbb{Z}_p$ is chosen to be surjective, \eqref{surfdefpart} is identical to the computation from the chain-saw quiver \cite{kantachi}. Note that the instanton part of the surface defect partition function also defines a statistical model on the set of colored partitions $\{ \boldsymbol{\lambda} \}$, with the measure $\boldsymbol{\mu}_{\boldsymbol{\lambda}}^{\mathbb{Z}_p, c} (\mathbf{a} , \boldsymbol{\varepsilon}) = \prod_{\omega \in \mathbb{Z}_p}  \mathfrak{q}_\omega ^{k_\omega} \epsilon ( - \mathcal{T}[\boldsymbol{\lambda}]^{\mathbb{Z}_p, c} )$.

\subsubsection{Degenerate gauge vertex construction} \label{deggauvert}
\begin{figure}[t]
\centering
\includegraphics[scale=0.4]{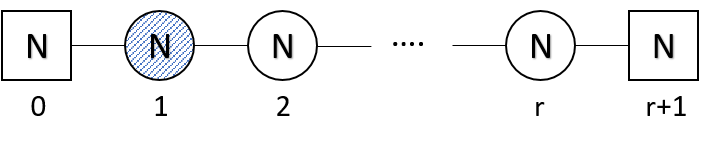}
\caption{$A_r$-quiver with a degenerate gauge vertex}
\label{fig1}
\end{figure}

For the $A_r$-quiver gauge theory with the gauge group $SU(N)$, the first gauge factor couples to a fundamental hypermultiplet which contributes to the partition function with its masses (See Figure \ref{fig1}). Let $a_{1,\alpha}$ be the Coulomb moduli for the first gauge vertex as before, and $a_{0,\alpha}$ be the mass of the hypermultiplet that couples to the first gauge vertex. If we tune the mass of the hypermultiplet as
\begin{align}
a_{0,1} &= a_{1,1} + \varepsilon_2 \nonumber \\
a_{0, \alpha} &= a_{1, \alpha} \quad \alpha \neq 1, \label{degcon}
\end{align}
it is apparant from \eqref{charac} that the contributions to the instanton partition function from all the fixed points vanish except the ones from single-column Young diagrams \cite{nek7},
\begin{align}
\boldsymbol{\lambda}^{(1)} = \left( \begin{rcases} \ytableausetup
{mathmode, boxsize=1em, centertableaux}
\begin{ytableau}
 \\ \\ \\ \none [\vdots] \\ \\
\end{ytableau} \end{rcases} d , \varnothing, \cdots, \varnothing  \right) \equiv \boldsymbol{\lambda}^{(1)} (d), \quad d \geq 0. \label{singcol}
\end{align}
Let us call a gauge vertex \textit{degenerate} in this case. In the BPS/CFT correspondence, the degenerate gauge vertex corresponds to inserting a simplest fully-degenerate primary field in $(r+3)$-point chiral block of $A_{N-1}$-Toda CFT\footnote{If we impose $a_{0,1} = a_{1,1} + \varepsilon_1$, it corresponds to the other simplest degenerate field in the $A_{N-1}$-Toda CFT and the surface defect on the other orthogonal plane. Here we choose to insert the degenerate vertex of type \eqref{degcon} to be consistent with the convention for the Nekrasov-Shatashvili limit, $\varepsilon_2 \to 0$, as we shall see.}, which was identified in \cite{agt2} with a half-BPS surface defect in the gauge theory. Hence the surface defect partition function is just the Nekrasov partition function \eqref{instpart} of the $A_r$-quiver gauge theory under the degenerate condition \eqref{degcon}. Now the $qq$-characters play an interesting role in connecting the gauge theory, the CFT, and the integrable system. The non-perturbative Dyson-Schwinger equations of the $A_r$-quiver gauge theory can be used to derive the BPZ equation with a simplest fully-degenerate field \cite{nek8}. The Nekrasov-Shatashvili limit (the semi-classical limit of the CFT) of the equation yields the Fourier transform of the Baxter equation for the corresponding integrable system. We elaborate on the derivation for the non-conformal $A_2$-theory with $SU(3)$ gauge group in section \ref{a2deggau}.

\subsection{Consequences of the non-perturbative Dyson-Schwinger equations} \label{sec3}
We now derive the differential equations that surface defect partition functions satisfy, using the non-perturbative Dyson-Schwinger equations. For generic quiver gauge theories with half-BPS surface defects, the non-perturbative Dyson-Schwinger equations derived in \cite{nek7} can be used to prove the KZ equation and the BPZ equation satisfied by the partition functions \cite{nek8}. In this paper, we study the surface defects on the pure $U(N)$ gauge theory which is relevant to the periodic Toda system. The orbifold surface defect partition function is shown to satisfy the Schr\"{o}dinger-type equation, while the degenerate gauge vertex partition function satisfies the Baxter-type equation. Note that those differential equations are valid for all values of $\boldsymbol{\varepsilon}=(\varepsilon_1, \varepsilon_2)$, as the fact will be crucial for investigating the special locus of the Coulomb moduli.
\subsubsection{$A_1$-theory with orbifold surface defect} \label{a1orb}
Let us consider the $A_1$-theory with the gauge group $U(N)$  in the presence of the regular orbifold surface defect $\mathcal{D}_{\mathbb{Z}_N, \rho}$, with the coloring function $s \in S_N$. With respect to the representations of $\mathbb{Z}_N$, the $\EuScript{Y}$-observable factors as:
\begin{align}
\EuScript{Y}(x)= \prod_{\omega \in \mathbb{Z}_N} \EuScript{Y}_\omega (x),
\end{align}
where
\begin{align}
\EuScript{Y}_\omega (x)[\boldsymbol{\lambda}] = (x - a_{s^{-1}(\omega)} ) \prod_{\Box \in K_\omega} \frac{x-c_{\hat{\Box}} -\varepsilon_1}{x-c_{\hat{\Box}}} \prod_{\Box \in K_{\omega-1} } \frac{x-c_{\hat{\Box}} - \varepsilon_2}{x-c_{\hat{\Box}}-\varepsilon}. \label{refy}
\end{align}
In terms of these $\EuScript{Y}_\omega$'s we also have the fundamental refined $qq$-characters, which are obtained as the orbifolded crossed instanton partition functions \cite{nek4},
\begin{align}
\EuScript{X}_{\omega}(x) = \EuScript{Y}_{\omega+1} (x+ \varepsilon) + \frac{\Lambda^2 z_\omega z_{\omega-1} ^{-1}} {\EuScript{Y}_\omega(x)},
\end{align}
whose expectation value in the gauge theory in the presence of the surface defect,
\begin{align}
\langle \EuScript{X}_\omega (x) \rangle_s \equiv \frac{1}{\boldsymbol{\Psi}_s ^{\text{inst}}}  \sum_{\boldsymbol{\lambda}} \EuScript{X}_\omega (\EuScript{Y}[\boldsymbol{\lambda}]) \mathfrak{q}^{\vert \boldsymbol{\lambda} \vert} \boldsymbol{\mu}^{\mathbb{Z}_N, s} _{\boldsymbol{\lambda}} (\mathbf{a},\boldsymbol{\varepsilon}) = T_{s, \omega}  (x), \label{expfundrefchar}
\end{align}
is a polynomial in $x$ by the compactness theorem proven in \cite{nek3}. In particular, we have the vanishing equations,
\begin{align}
[x^{-n}] \langle \EuScript{X}_\omega (x) \rangle_s = 0, \quad n\in \mathbb{Z}_{>0}.
\end{align}
We study the coefficients of $x^{-n}$ of the fundamental refined $qq$-character in the large $x$ limit. The lowest order coefficients are given by:
\begin{subequations} \label{coeff}
\begin{align}
[x^{-1}]\EuScript{X}_\omega &=  \frac{\varepsilon_1 ^2}{2} \left( k_\omega - k_{\omega +1} - \frac{a_{s^{-1}(\omega+1)}}{\varepsilon_1} \right)^2 - \frac{1}{2} a_{s^{-1} (\omega +1)} ^2 + \varepsilon_1 \varepsilon_2 k_{\omega} + \Lambda^2 z_\omega z_{\omega-1} ^{-1}  \nonumber \\
& \quad \quad \quad  + \frac{\varepsilon_1^2}{2} (k_\omega - k_{\omega+1}) + \varepsilon_1 \left(\sum_{\Box \in K_{\omega}} c_{\hat{\Box}} - \sum_{\Box \in K_{\omega+1}} c_{\hat{\Box}} \right) , \label{coeff-2} \\ 
[x^{-2}]\EuScript{X}_\omega &= \frac{\varepsilon_1 ^3}{6}(k_\omega -k_{\omega+1})^3 - \frac{\varepsilon_1 ^3}{2}(k_\omega - k_{\omega+1})^2 + \varepsilon_1 ^2 \varepsilon_2 k_{\omega+1} (k_\omega - k_{\omega+1}) \nonumber \\
& + (\varepsilon -a_{s^{-1}(\omega+1)} )\left( \frac{\varepsilon_1 ^2}{2} (k_\omega - k_{\omega+1})^2 - \frac{\varepsilon_1 ^2}{2} (k_\omega - k_{\omega+1}) + \varepsilon_1 \varepsilon_2 k_{\omega+1} + \varepsilon_1 \left( \sum_{\Box \in K_\omega} c_{\hat{\Box}} - \sum_{\Box \in K_{\omega+1}} c_{\hat{\Box}} \right) \right) \nonumber \\
& + \Lambda^2 z_\omega z_{\omega-1} ^{-1} (a_{s^{-1}(\omega)} + \varepsilon_1 (k_\omega - k_{\omega-1})) + \varepsilon_1 ^2 (k_\omega - k_{\omega+1}) \left(  \sum_{\Box \in K_\omega} c_{\hat{\Box}} - \sum_{\Box \in K_{\omega+1}} c_{\hat{\Box}} \right) \nonumber \\
& + \frac{\varepsilon_1 ^3}{3} (k_\omega - k_{\omega+1}) - \varepsilon_1 ^2 \left( \sum_{\Box \in K_\omega} c_{\hat{\Box}} - \sum_{\Box \in K_{\omega+1}} c_{\hat{\Box}} \right) + \varepsilon_1 \left(  \sum_{\Box \in K_\omega} c_{\hat{\Box}} ^2 - \sum_{\Box \in K_{\omega+1}} c_{\hat{\Box}} ^2 \right) \nonumber \\
& - \varepsilon_1 \varepsilon_2 \varepsilon k_{\omega+1} + 2 \varepsilon_1 \varepsilon_2 \sum_{\Box \in K_{\omega+1}} c_{\hat{\Box}}. \label{coeff-3}
\end{align}
\end{subequations}
The expectation values of \eqref{coeff} yield the vanishing equations. We take the sum over $\omega \in \mathbb{Z}_N$, while simplifying \eqref{coeff-3} using \eqref{coeff-2}, to get the following differential equations,
\begin{subequations} \label{schrodingers}
\begin{align} 0 &= \left[ \frac{\varepsilon_1 ^2}{2} \sum_\omega  \left( z_\omega \frac{\partial}{\partial z_\omega} - \frac{a_{s^{-1}(\omega+1)}}{\varepsilon_1} \right)^2 + \Lambda^2 \sum_\omega z_\omega z_{\omega -1} ^{-1} - \frac{1}{2} \sum_\omega a_{s^{-1}(\omega+1)} ^2  + \frac{1}{2} \varepsilon_1 \varepsilon_2 \Lambda \frac{\partial}{\partial \Lambda} \right] \boldsymbol{\Psi}^{\text{inst}}_s(\mathbf{a},\boldsymbol{\varepsilon},\mathfrak{q}, \mathbf{z}), \label{schrodinger} \\
0 &= \left[ - \frac{\varepsilon_1 ^3}{3} \sum_\omega \left( z_\omega \frac{\partial}{\partial z_\omega} - \frac{a_{s^{-1}(\omega+1)}}{\varepsilon_1} \right)^3 \right.  \nonumber \\
& \quad\quad + \Lambda^2 \sum_\omega z_\omega z_{\omega-1} ^{-1} \left( -\varepsilon_1 \left(z_\omega \frac{\partial}{\partial z_\omega} + z_{\omega-1} \frac{\partial}{\partial z_{\omega-1}} - \frac{a_{s^{-1}(\omega+1)}}{\varepsilon_1} - \frac{a_{s^{-1}(\omega)}}{\varepsilon_1} \right) + \varepsilon_2  \right) \nonumber \\
& \quad\quad \left. -\frac{1}{3} \sum_\omega a_{s^{-1}(\omega+1)} ^3 + \frac{1}{2} \varepsilon_1 \varepsilon_2 \varepsilon \Lambda \frac{\partial}{\partial \Lambda} + 2\varepsilon_1 \varepsilon_2 \Bigg\langle \sum_{\Box \in K} c_{\hat{\Box}} \Bigg\rangle_s \right] \boldsymbol{\Psi}^{\text{inst}}_s(\mathbf{a},\boldsymbol{\varepsilon},\mathfrak{q}, \mathbf{z}). \label{schrodinger3}
\end{align}
\end{subequations}
Note that \eqref{schrodinger} is the one-line rederivation of the results of \cite{bra,braeti}. In the Nekrasov-Shatashvili limit ($\varepsilon_2 \rightarrow 0$), these differential equations produce the spectral equations for the Hamiltonians $\EuScript{O}_2$ and $\EuScript{O}_3$ of the periodic Toda system, as we shall see shortly in section \ref{splitting}.

\subsubsection{$A_2$-theory with degenerate gauge vertex} \label{a2deggau}
The fundamental $qq$-characters for the $A_2$-quiver gauge theory can be written as \cite{nek2,nek4},
\begin{subequations} \label{a2qq}
\begin{align}
\EuScript{X}_1 (x) &= \EuScript{Y}_1 (x + \varepsilon) + \mathfrak{q}_1 \frac{\EuScript{Y}_0 (x) \EuScript{Y}_2 (x+\varepsilon)}{\EuScript{Y}_1 (x)} + \mathfrak{q}_1 \mathfrak{q}_2 \frac{\EuScript{Y}_0 (x) \EuScript{Y}_3 (x+ \varepsilon) }{\EuScript{Y}_2 (x)}, \\
\EuScript{X}_2 (x) &= \EuScript{Y}_2 (x+ \varepsilon) + \mathfrak{q}_2 \frac{\EuScript{Y}_1 (x) \EuScript{Y}_3 (x+\varepsilon)}{\EuScript{Y}_2 (x)} + \mathfrak{q}_1 \mathfrak{q}_2 \frac{\EuScript{Y}_0 (x-\varepsilon) \EuScript{Y}_3 (x+ \varepsilon)}{\EuScript{Y}_1 (x - \varepsilon)}, \label{a2qqchar}
\end{align}
\end{subequations}
\begin{figure}[b]
\centering
\includegraphics[scale=0.4]{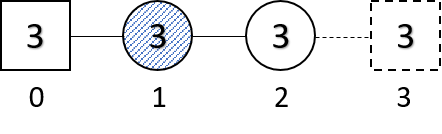}
\caption{non-conformal $A_2$-quiver with a degenerate gauge vertex, $SU(3)$ gauge group}
\label{fig2}
\end{figure}whose expectation values are polynomials $\langle \EuScript{X}_i (x) \rangle = T_i (x)$ \cite{nek3}. Here $\EuScript{Y}_0 (x) = \prod_{\alpha=1} ^N (x-a_{0,\alpha} )$ and $\EuScript{Y}_3 (x) = \prod_{\alpha=1} ^N (x-a_{3,\alpha})$ by definition, $\EuScript{Y}_{1,2} (x)$ are $\EuScript{Y}$-observables \eqref{yobsst} for the two gauge vertices, and $\mathfrak{q}_{1,2}$ are the respective gauge couplings. Note that some of the terms are simplified if we start from the non-conformal $A_2$-theory with the gauge group $SU(3)$, in which the second hypermultiplet decouples
\begin{align}
\mathfrak{q}_2 \rightarrow 0, \quad a_{3,\alpha} \rightarrow \infty, \quad \mathfrak{q}_2 \prod_{\alpha=1} ^3 a_{3,\alpha} = \Lambda^3 \text{     fixed}.
\end{align}
In the $A_2$-Toda CFT side, this is equivalent to studying the irregular 4-point block (one irregular puncture, one full puncture, and two semi-degenerate punctures). Now we replace one of the semi-degenerate fields by a simplest fully-degenerate field, making the degenerate irregular 4-point block. As we have seen in section \ref{deggauvert} this replacement corresponds to degenerate gauge vertex, which we choose to be the left-most one (See Figure \ref{fig2}). When the first gauge vertex is degenerate in the sense of \eqref{degcon}, it contributes to the instanton partition function only through the single-column Young diagrams \eqref{singcol}, and the first $\EuScript{Y}$-observable is simplified,
\begin{align}
\EuScript{Y}_1(x)[\boldsymbol{\lambda}^{(1)} (d)] = \EuScript{Y}_0 (x) \frac{x-a_{0,1}+\varepsilon_2 - d \varepsilon_1}{x-a_{0,1}- d \varepsilon_1}. \label{a1y1}
\end{align}
Then the two $qq$-character equations \eqref{a2qq} can be combined to yield a degree 4 equation
\begin{align}
0 &= \left(x-a_{0,1} - \varepsilon_1 \mathfrak{q}_1 \frac{\partial}{\partial \mathfrak{q}_1} \right)(T_2 (x) \EuScript{Z}) + \mathfrak{q}_1 \Lambda^3 \left(x-a_{0,1} - \varepsilon_2 - \varepsilon_1 \left( 1+ \mathfrak{q}_1 \frac{\partial}{\partial \mathfrak{q}_1} \right) \right) \EuScript{Z} \nonumber \\
& \quad\quad -\mathfrak{q}_1 ^{-1}  \left(x-a_{0,1}+ \varepsilon_2 - \varepsilon_1 \left(-1 + \mathfrak{q}_1  \frac{\partial}{\partial \mathfrak{q}_1 } \right) \right) (T_1 (x) \EuScript{Z}) \nonumber \\
&\quad \quad + \mathfrak{q}_1 ^{-1} \EuScript{Y}_0 (x+ \varepsilon) \left( x - a_{0,1} + 2 \varepsilon_2 - \varepsilon_1 \left( -1 + \mathfrak{q}_1  \frac{\partial}{\partial \mathfrak{q}_1 } \right) \right) \EuScript{Z}.
\end{align}
All the coefficients of positive powers of $x$ actually yield trivial equations. The coefficient of $x^0$ gives the following differential equation for the partition function with respect to $\mathfrak{q}_1$ and $\Lambda$,
\begin{align}
0=&\left[- \mathfrak{q}_1^{-1} \prod_{\alpha=1} ^3 \left(a_{0,1}-a_{0,\alpha} + \varepsilon_1 \mathfrak{q}_1 \frac{\partial}{\partial \mathfrak{q}_1} \right) + \prod_{\alpha=1} ^3 \left( a_{0,1} -a_{2,\alpha} + \varepsilon_1 \left( 1+ \mathfrak{q}_1 \frac{\partial}{\partial \mathfrak{q}_1} \right) \right) \right. \nonumber \\
&\left. \quad + \varepsilon_1 \varepsilon_2 \left(\varepsilon + \varepsilon_1 +a_{0,1} + \varepsilon_1 \mathfrak{q}_1 \frac{\partial}{\partial \mathfrak{q}_1} \right) \frac{1}{3} \Lambda \frac{\partial}{\partial \Lambda} - \Lambda^3 (1-\mathfrak{q}_1) + 2\varepsilon_1 \varepsilon_2\  \Bigg\langle  \sum_{\Box \in K_2} c_{\hat{\Box}} \Bigg\rangle  \right] \EuScript{Z}(\mathbf{a},\boldsymbol{\varepsilon},\mathfrak{q}_1, \Lambda) .\label{BPZ}
\end{align}
In the context of the BPS/CFT correspondence, this is the BPZ equation \cite{bpz} for the degenerate irregular 4-point block in the $A_2$-Toda CFT. Note that it is valid in the full quantum regime of the CFT.

For our purpose of investigating the periodic Toda system, we can subject \eqref{BPZ} to a further degeneration limit. First we make a change of variables,
\begin{align}
\mathfrak{q}_1 \mapsto z(\mathfrak{q}_1), \quad \frac{d \text{log} (z(\mathfrak{q}_1))}{d \mathfrak{q}_1}&=\mathfrak{q}_1 ^{-2/3} (\mathfrak{q}_1 -1)^{-1/3},
\end{align}
introduce a prefactor,
\begin{align}
\EuScript{Z}(\mathbf{a}, \boldsymbol{\varepsilon}, \mathfrak{q}_1, \Lambda) &= e^{f(\mathfrak{q}_1)}\widetilde{\EuScript{Z}}(\mathbf{a}, \boldsymbol{\varepsilon}, z, \widetilde{\Lambda}), \\ 
\text{with} \quad \frac{d f(\mathfrak{q}_1)}{d \mathfrak{q}_1} &= \frac{-2a_{0,1} +a_{0,2}+a_{0,3}-\varepsilon_1 + 3(a_{0,1}+\varepsilon_1)\mathfrak{q}_1}{3 \varepsilon_1 \mathfrak{q}_1 ^{1/3} (\mathfrak{q}_1 -1)^{2/3}}, \nonumber
\end{align}
to bring the equation into the canonical form in which the coefficients of the highest order differential is 1 and of the second-highest order differential is 0. Then we get the degenerate irregular 3-point block (two irregular punctures and one fully-degenerate puncture) by taking the limit
\begin{align}
&\Lambda \rightarrow 0, \quad a_{0,\alpha} \rightarrow \infty, \quad \mathfrak{q}_1 \rightarrow \infty, \nonumber \\
&\Lambda^3 \prod_{\alpha=1} ^3 a_{0,\alpha} = \widetilde{\Lambda}^6, \quad \frac{\mathfrak{q}_1}{\prod_{\alpha=1} ^3 a_{0,\alpha}} = z \quad \text{fixed}.
\end{align} 
Even though the result is fairly complicated in general case, only a few terms survive in the degeneration limit. The final result is
\begin{align}
0 = &\left[ \widetilde{\Lambda}^6 z + \frac{1}{z}  + \varepsilon_1 ^3 \left( z \frac{\partial}{\partial z} \right)^3 - \frac{\varepsilon_1}{2} \left( \sum_{\alpha=1} ^3 a_{2,\alpha} ^2  - \frac{1}{3} \varepsilon_1 \varepsilon_2 \widetilde{\Lambda} \frac{\partial}{\partial \widetilde{\Lambda}}\right)\left(z \frac{\partial}{\partial z} \right) \right. \nonumber \\
&\left. \quad \quad \quad \quad \quad \quad \quad \quad  - \frac{1}{3} \sum_{\alpha=1} ^3 a_{2,\alpha} ^3 + \frac{1}{6}\varepsilon_1 \varepsilon_2 \varepsilon \widetilde{\Lambda} \frac{\partial}{\partial \widetilde{\Lambda}} + 2 \varepsilon_1 \varepsilon_2 \Bigg\langle \sum_{\Box \in K_2} c_{\hat{\Box}} \Bigg\rangle  \right]\widetilde{\EuScript{Z}}(\mathbf{a}, \boldsymbol{\varepsilon}, z, \widetilde{\Lambda}). \label{BPZ'}
\end{align}
Thus we have derived the BPZ equation for the degenerate irregular 3-point block in the $A_2$-Toda CFT. Note that this equation can be interpreted as the double quantization of the Seiberg-Witten curve \eqref{speccurve} for the pure $\EuScript{N}=2$ $SU(3)$ gauge theory. Under the Fourier transform
\begin{align}
\widetilde{\EuScript{Z}}(\mathbf{a}, \boldsymbol{\varepsilon}, z, \widetilde{\Lambda}) \equiv \sum_{x \in \Gamma} \EuScript{Q}(x) z^{-\frac{x}{\varepsilon_1}},
\end{align}
where $\Gamma$ is a lattice with the lattice spacing $\varepsilon_1$, the equation becomes the Baxter-type equation
\begin{align}
\widetilde{\Lambda}^6 \EuScript{Q} (x+\varepsilon_1) +  \EuScript{Q}(x-\varepsilon_1) = T(x) \EuScript{Q}(x), \label{qbax}
\end{align}
where we have defined the spectral polynomial
\begin{align}
T(x) = x^3 - \frac{1}{2} \langle \EuScript{O}_2 \rangle x + \frac{1}{3} \langle \EuScript{O}_3 \rangle, \label{transfer}
\end{align}
whose coefficients are precisely the expectation values of the gauge-invariant chiral observables \eqref{obsst}. Note that \eqref{qbax} indeed has the form of the Baxter equation for the periodic Toda system\footnote{Since we have started with the gauge group $SU(3)$ instead of $U(3)$, the total momentum $\EuScript{O}_1$ has been decoupled.} \cite{gp}, except it is more general since it is valid for generic values of $\boldsymbol{\varepsilon} = (\varepsilon_1 , \varepsilon_2)$. In the Nekrasov-Shatashvili limit $\varepsilon_2 \to 0$ (the semi-classical limit of the CFT, $\hbar_{\text{CFT}} ^2 = \varepsilon_1 \varepsilon_2 \to 0$), the equation \eqref{qbax} is obviously reduced to the ordinary Baxter equation for the 3-particle periodic Toda system.

\section{Splitting of the surface defect partition function} \label{splitting}
Finally we study the splitting behavior of the regular orbifold surface defect partition functions and its relation with integrable systems. A crucial remark is that the differential equations \eqref{schrodingers} are still valid even at the special locus of the Coulomb moduli, $\left\{\frac{a_{\alpha \beta}}{\varepsilon_1} \in \mathbb{Z} \setminus \{0\} \right\}$. Thus the surface defect partition function can be used as a probe for the special locus, where the bulk partition function does not provide a simple picture for the correspondence. Meanwhile, on the integrable system side the special locus still gives the well-defined spectral problem of mutually commuting Hamiltonians, except that the spectra become degenerate at the 0-th order due to the specially tuned Bloch angles. In particular, the differential equations that define the spectral problem are still the same. Therefore the surface defect partition function is expected to detect such a splitting behavior of the corresponding integrable system. In particular, we will observe that, while the surface defect partition function still has the additional singularities in the limit $\varepsilon_2 \to 0$, it splits into parts in such a way that those extra singularities are resolved in each split part.

First note that for generic values of Coulomb moduli the surface defect partition function exhibits the typical asymptotic behavior in $\varepsilon_2 \rightarrow 0$,
\begin{align}
\widetilde{\boldsymbol{\Psi}}_s(\mathbf{a},\boldsymbol{\varepsilon},\Lambda, \mathbf{z}) \equiv \prod_\omega z_\omega ^{-\frac{a_{s^{-1}(\omega+1)}}{\varepsilon_1}}  \boldsymbol{\Psi}^{\text{inst}}_s(\mathbf{a},\boldsymbol{\varepsilon},\Lambda, \mathbf{z}) = e^{\frac{\widetilde{\EuScript{W}}( \mathbf{a},\varepsilon_1, \Lambda)}{\varepsilon_2}} (\psi_s (\mathbf{a},\varepsilon_1, \Lambda,\mathbf{z}) +\mathcal{O}(\varepsilon_2)), \label{asymp}
\end{align}
up to some prefactor. Therefore the differential equations \eqref{schrodingers} realize the Schr\"{o}dinger equations for the periodic Toda system
\begin{subequations} \label{schrodingers'}
\begin{align}
&\left[ \frac{\varepsilon_1 ^2}{2} \sum_\omega  \left( z_\omega \frac{\partial}{\partial z_\omega} \right)^2 + \Lambda^2 \sum_\omega z_\omega z_{\omega-1}^{-1} - E_2(\mathbf{a},\varepsilon_1, \Lambda) \right] \psi_s (\mathbf{a},\varepsilon_1, \Lambda, \mathbf{z})  = 0, \label{schrodinger'} \\
&\left [ - \frac{\varepsilon_1 ^3}{3} \sum_\omega \left( z_\omega \frac{\partial}{\partial z_\omega} \right)^3 -\varepsilon_1 \Lambda^2 \sum_\omega z_\omega z_{\omega-1} ^{-1} \left( z_\omega \frac{\partial}{\partial z_\omega} +z_{\omega-1} \frac{\partial}{\partial z_{\omega-1}} \right) - E_3(\mathbf{a},\varepsilon_1, \Lambda) \right] \psi_s (\mathbf{a},\varepsilon_1, \Lambda, \mathbf{z})  = 0, \label{schrodinger3'}
\end{align}
\end{subequations}
where
\begin{subequations} \label{energies'}
\begin{align}
E_2(\mathbf{a},\varepsilon_1, \Lambda) &= \frac{1}{2} \sum_\omega a_\omega ^2 - \frac{1}{2} \varepsilon_1 \Lambda \frac{\partial \widetilde{\EuScript{W}}(  \mathbf{a},\varepsilon_1, \Lambda)}{\partial \Lambda}, \label{energy'} \\
E_3(\mathbf{a},\varepsilon_1, \Lambda) &= \frac{1}{3} \sum_\omega a_\omega ^3 - \frac{1}{2} \varepsilon_1 ^2 \Lambda \frac{\partial \widetilde{\EuScript{W}}(  \mathbf{a},\varepsilon_1, \Lambda)}{\partial \Lambda} - 2 \varepsilon_1 \lim_{\varepsilon_2 \to 0} \varepsilon_2 \Bigg\langle \sum_{\Box \in K} c_{\hat{\Box}} \Bigg\rangle_s \label{energy3'}
\end{align}
\end{subequations}
are nothing but the eigenvalues of the Hamiltonians \eqref{energies} we have derived in the theory without the surface defect.\footnote{The relative factor $N$ in the second term is due to the map $(z_1 , z_2) \mapsto (z_1 , z_2 ^N)$ in the orbifold construction of the regular orbifold surface defect, which shifts the equivariant parameter as $\varepsilon_2 \rightarrow N\varepsilon_2$.}\footnote{Although the eigenvalue \eqref{energy3'} seems to depend on the choice $s \in S_N$ through the expectation value $\langle \cdots \rangle_s$, it turns out not to. This is consistent with the computation in the absence of the surface defect, \eqref{energy3}.} Note that even though the meaning of the expectation values in \eqref{energy3} and \eqref{energy3'} are different, the final results agree in the limit $\varepsilon_2 \to 0$. Thus the surface defect partition function provides a constructive way to obtain both the eigenfunctions and the eigenvalues of the Hamiltonians of the corresponding integrable system.

Now we attempt an analogous construction at the special locus of the Coulomb moduli. The investigation reveals the splitting behavior of the surface defect partition functions.

\subsection{$N=2$}
Let us first consider the simplest case, $N=2$, in which there are two choices for the regular orbifold surface defect corresponding to the elements of $S_2 = \{ \text{id}, (01) \}$. The Schr\"{o}dinger equation \eqref{schrodinger'} is precisely the Mathieu equation up to some change of variables. At the special locus $ \{ a_{01} = m \varepsilon_1 \; \vert \; m \in \mathbb{Z} \setminus \{0\} \} $, we observe that the surface defect partition functions split into two parts,
\begin{align}
\widetilde{\boldsymbol{\Psi}}_{\text{id}} (a_{01} =m\varepsilon_1, \boldsymbol{\varepsilon}, \Lambda, \mathbf{z}) \pm \widetilde{\boldsymbol{\Psi}}_{(01)} (a_{01} =m\varepsilon_1 ,\boldsymbol{\varepsilon}, \Lambda, \mathbf{z}) = e^{\frac{\widetilde{\EuScript{W}}_m ^\pm ( \varepsilon_1, \Lambda)}{\varepsilon_2}} \left(\psi_m ^\pm (\varepsilon_1, \Lambda, \mathbf{z}) + \mathcal{O}(\varepsilon_2)\right). \label{split2}
\end{align}
Note that \eqref{schrodinger} guarantees the wavefunctions $\psi_m ^\pm (\varepsilon_1, \Lambda, \mathbf{z} )$ to be the split eigenfunctions of the Schr\"{o}dinger equation \eqref{schrodinger'} with the split energy spectrum
\begin{align}
E_{2, m}^\pm = \frac{m^2 \varepsilon_1 ^2}{8} -\frac{1}{4} \varepsilon_1 \Lambda \frac{\partial \widetilde{\EuScript{W}}_m ^\pm (\varepsilon_1, \Lambda)}{\partial \Lambda}. \label{energydeg}
\end{align}
We decoupled the irrelevant center of mass contribution and rescaled by a factor of 2 for convenience. The splitting behavior exactly accounts for the broken degeneracy due to the quantum tunneling effects on the integrable system side. Note that \eqref{split2} is not obvious in the sense that the \textit{split twisted superpotential} $\widetilde{\EuScript{W}}_m ^\pm$ is non-divergent and is independent of the fractional gauge coupling $\mathbf{z}$. Also, it should be emphasized that the split twisted superpotential $\widetilde{\EuScript{W}}_m ^\pm$ shows the series expansion in $\Lambda^2$, as opposed to the $\Lambda^4$-expansion of the generic twisted superpotential. 

We have checked that the split eigenfunctions $\psi_m ^\pm$ and the split eigenvalues $E_{2,m} ^\pm$ in \eqref{split2} and \eqref{energydeg} precisely match with the well-known results of the half-periodic and the periodic solutions for the Mathieu equation, for various $m \in \mathbb{Z} \setminus \{0\}$ to some order of $\Lambda$. Therefore the splitting of the surface defect partition functions accounts for the splitting of the degenerate levels in the integrable system, and the correspondence between the gauge theory and the integrable system is recovered for the special locus of the Coulomb moduli space. We present some specific examples of the computation in Appendix \ref{examcomp}.

\subsection{$N=3$}
In the case $N=3$, the Hamiltonians are no longer Hermitian and the eigenvalues are not necessarily real, yet the perturbative series is well-defined including the degenerate case. Therefore we can still compare the spectra and the wavefunctions obtained from the gauge theory with the quantum mechanical computations. As mentioned in section \ref{bethegauge}, for the non-degenerate cases the known dictionary of the correspondence works as stated. Let us turn to the degenerate cases. There are three types of degeneracy possible, which are 2-fold, 3-fold, and 6-fold respectively. Without loss of generality, those degeneracies occur at the loci
\begin{align}
\text{2-fold :}& \quad \{ a_{01} = m \varepsilon_1, a_{02} \text{ is generic} \; \vert \; m \in \mathbb{Z} \setminus \{0\} \} \nonumber \\
\text{3-fold :}& \quad \{ a_{01} =a_{02} = m\varepsilon_1 \; \vert \; m \in \mathbb{Z} \setminus \{0\} \} \nonumber \\
\text{6-fold :}& \quad \{ a_{01} = m \varepsilon_1, a_{02} = l \varepsilon_1 \; \vert \; m, l \in \mathbb{Z} \setminus \{0\} , m \neq l \}. \nonumber
\end{align}
There are some subtle issues for the 2-fold and 6-fold degeneracies that obstruct our understanding of the splitting of the surface defect partition function, so we leave them to future work. Here we discuss the splitting of the surface defect partition function for the 3-fold degeneracy.

We have 6 different regular surface defects corresponding to the elements $s \in S_3$. Due to the residual symmetry, only 3 out of 6 are independent of each another in the case of $a_{12}=0$. We form the split surface defect partition functions as
\begin{align}
&\left[ \widetilde{\boldsymbol{\Psi}}_\text{(012)} (\boldsymbol{a} ,\boldsymbol{\varepsilon}, \Lambda, \mathbf{z}) + \zeta \widetilde{\boldsymbol{\Psi}}_\text{(021)} (\boldsymbol{a} ,\boldsymbol{\varepsilon}, \Lambda, \mathbf{z}) + \zeta^2 \widetilde{\boldsymbol{\Psi}}_\text{id} (\boldsymbol{a},\boldsymbol{\varepsilon}, \Lambda, \mathbf{z}) \right] \Bigg\vert_{a_{01} = a_{02} = m\varepsilon_1} \quad \quad \quad \quad\quad \quad\nonumber \\
&\quad \quad\quad \quad\quad \quad\quad \quad\quad \quad\quad \quad\quad \quad\quad \quad\quad \quad\quad \quad= e^{ \frac{\widetilde{\EuScript{W}}_m ^\zeta (\varepsilon_1, \Lambda)}{\varepsilon_2}} \left(\psi_m ^\zeta (\varepsilon_1, \Lambda, \mathbf{z}) + \mathcal{O}(\varepsilon_2)\right),
\end{align}
where $\zeta$ is any third root of unity, $\zeta^3 =1$. Therefore each surface defect partition function splits into three parts, accounting for the level splitting of the 3-fold degeneracy. The wavefunctions $\psi_m ^\zeta (\varepsilon_1, \Lambda, \mathbf{z})$ are the common split eigenfunctions of $\EuScript{O}_2$ and $\EuScript{O}_3$ by \eqref{schrodingers} with the split eigenvalues
\begin{subequations} \label{dics}
\begin{align}
E_{2,m} ^\zeta &= \frac{m^2 \varepsilon_1 ^2}{3} - \frac{1}{2} \varepsilon_1 \Lambda \frac{\partial \widetilde{\EuScript{W}}_m ^\zeta (\varepsilon_1, \Lambda)}{\partial \Lambda}, \label{dic2} \\
E_{3,m} ^\zeta &= \frac{2 m^3 \varepsilon_1 ^3 }{27} -\frac{\varepsilon_1 ^2}{2} \Lambda \frac{\partial \widetilde{\EuScript{W}}_m ^\zeta (\varepsilon_1, \Lambda)}{\partial \Lambda} - 2 \varepsilon_1 c_1 ^\zeta (\varepsilon_1, \Lambda), \label{dic3}
\end{align}
\end{subequations}
where we have decoupled the irrelevant center of mass contribution and defned the \textit{split expectation value}
\begin{align}
&c_1 ^\zeta (\varepsilon_1, \Lambda) \nonumber \\
& =\lim_{\varepsilon_2 \to 0} \varepsilon_2 \frac{ \Bigg\langle \sum_{\Box \in K} c_{\hat{\Box}} \Bigg\rangle_{\text{(012)}} \widetilde{\Psi}_\text{(012)} + \zeta \Bigg\langle \sum_{\Box \in K} c_{\hat{\Box}} \Bigg\rangle_{\text{(021)}} \widetilde{\Psi}_\text{(021)} + \zeta^2 \Bigg\langle \sum_{\Box \in K} c_{\hat{\Box}} \Bigg\rangle_{\text{id}} \widetilde{\Psi}_\text{id} }{\widetilde{\Psi}_{\text{(012)}} +\zeta \widetilde{\Psi}_{\text{(021)}} +\zeta^2 \widetilde{\Psi}_{\text{id}} }  \Bigg\vert_{a_{01}=a_{02}=m \varepsilon_1}.
\end{align}
It is not obvious that $c_1 ^\zeta (\varepsilon_1, \Lambda)$ neither diverges nor depends on the fractional coupling $\mathbf{z}$; in those cases the split eigenvalue \eqref{dic3} would not be well-defined. The computation shows that $c_1 ^\zeta (\varepsilon_1, \Lambda)$ indeed behaves as desired. Note that the split twisted superpotential $\widetilde{\EuScript{W}}_m ^\zeta$ and the split expectation value $c_1 ^\zeta$ have the series expansions in $\Lambda^2$, as opposed to the $\Lambda^6$-expansion of the generic twisted superpotential and expectation value. We present some examples of computation in Appendix \ref{examcomp2}.

\section{Discussion} \label{discussion}
In this paper we have studied the Bethe/gauge correspondence for the special locus of the Coulomb moduli of the gauge theory, where the integrable system becomes degenerate in the non-interacting (free) limit. The analysis on the gauge theory with partial noncommutativity and partial $\Omega$-deformation revealed the emergence of extra massless modes of matter multiplet at the speical locus, which makes the generic effective description without matter multiplet inapplicable. We used half-BPS surface defects, which are constructed out of orbifold and degenerate gauge vertex, to investigate the problem. The orbifold surface defect provided a constructive approach for the common eigenfunctions as well as the spectra of the Hamiltonians of the integrable system. Namely, the non-perturbative Dyson-Schwinger equations can be used to show that the surface defect partition function satisfies the Schr\"{o}dinger-type equations, which indeed reduce to the spectral equations for the Hamiltonians in the Nekrasov-Shatashvili limit. The degenerate gauge vertex partition function was shown to satisfy the BPZ equation of the dual CFT by the non-perturbative Dyson-Schwinger equation. In the Nekrasov-Shatashvili limit, the equation was reduced to the Fourier transform of the Baxter equation for the corresponding integrable system. We have seen that at the special locus of the Coulomb moduli the orbifold surface defect partition functions split into parts. Each split part assumes the desired asymptotic behavior in the Nekrasov-Shatashvili limit so that the degenerate perturbative series for the eigenfunctions and the eigenvalues could be presicely reproduced from the gauge theory perspective. We have presented some examples of the splitting.

There is a natural generalization of the investigation, i.e. adding various flavors to the theory. It is manifest from the instanton counting procedure that the theories with various types of flavor share the same denomenator in the effective twisted superpotential. Thus $U(N)$ gauge theories with flavors show the same divergent phenomena at the special locus of the Coulomb moduli space, which are expected to correspond to the splitting of degeneracies in the integrable system side. We may introduce the regular surface defect in those theories, with some proper assignment of the colorings for the flavors, and investigate the splitting behavior at the special locus. Some theories with fundamental hypermultiplets have non-Hermitian Hamiltonians even in the simplest case $N=2$. It would be a nontrivial check to see how the splitting works for those theories.

Another interesting issue to be considered is the 5d uplift. While $d=4$, $\EuScript{N}=2$ gauge theories correspond to the non-relativistic integrable systems realized on the Seiberg-Witten geometry, the $d=5$, $\EuScript{N}=1$ gauge theories compactified on a circle correspond to their relativistic cousins \cite{nek9}. The main difference is that the spectral equations become difference equations instead of differential equations. It was checked in \cite{bkk} at some low instanton numbers that the codimension-two surface defect partition function satisfies those difference equations, for the example of $\EuScript{N}=1^*$ theory. It would be nice to construct a rigorous analytic proof of those relations as done in this work for the four-dimensional case, using the 5d version of the $qq$-characters \cite{nek2}. The \textit{algebraic engineering} of codimension-two defect partition functions \textit{\`{a} la} \cite{algen} can be useful for this study. The splitting of degeneracies would persist in those relativistic integrable systems, and the insertion of codimension-two defects is expected to detect this splitting through their partition functions.

The study of resurgence in integrable systems can have a connection with our story. For example, let us consider the Mathieu system which corresponds to the pure $\EuScript{N}=2$ $SU(2)$ gauge theory. The exact spectrum of the Mathieu system around a minimum of the Mathieu potential $V(x)=\Lambda^2 \text{cos}x$ exhibits the trans-series expansion, which can be computed by the exact quantization condition \cite{dunun}. In \cite{kre}, it was argued that this exact quantization condition can be regarded as the Nekrasov-Shatashvili quantization condition in the strong coupling regime. The analysis showed that the prepotential at the strong coupling regime gets non-perturbative corrections (in the sense of quantum mechanics). Using the connection between the weak and the strong coupling regimes described in \cite{gokdun}, we may look for the gauge theoretical understanding of a nontrivial relation between the aformentioned non-perturbative effect in the strong coupling regime and the non-perturbative effect in the weak coupling regime, i.e. the splitting of the degenerate levels studied in this paper. The topological string point of view on the exact quantization in \cite{wzh} can also be related along these lines.

It would also be interesting to clarify the implication of the other eigenfunctions for the split eigenvalues. For example, it is well-known that for the Mathieu system the second solution for the split eigenvalue includes a $\text{log}z$ term. Actually the second solution for $a_{01}=0$ (where the splitting does not occur) can be obtained by taking a derivative of the surface defect partition function with respect to the Coulomb moduli. When $a_{01} = m \varepsilon_1$ this procedure is not available since the surface defect partition function has discontinuity at the special locus. However, we may insert a 't Hooft line operator on top of the surface operator to get a $\varepsilon_2$-shift of the Coulomb moduli \cite{agt2, tes,pesoku}, which becomes infinitesimal in the Nekrasov-Shatashvili limit. Since the configuration is expected to have a well-defined effective twisted superpotential in the Nekrasov-Shatashvili limit, its partition function may produce the second solution with log. Unfortunately, the supersymmetric localization for such configuration of non-local observables is not available as of yet. 

\acknowledgments

The author is deeply benefited from Nikita Nekrasov for invaluable suggestions and discussions. The author also would like to thank Alex DiRe, Hee-Cheol Kim, Naveen Prabhakar, Antonio Sciarappa, Jaewon Song, and Xinyu Zhang for useful discussions on relevant subjects. The research was supported by the NSF grant PHY 1404446, and also by the Samsung Scholarship.

\appendix
\section{Examples of split surface defect partition functions}
\subsection{$N=2$}  \label{examcomp}
For $N=2$ case we can compare the results from the gauge theory with the well-known Mathieu functions with half-period and whole-period, in \cite{wang} for example. We observe the precise match between the two.

\paragraph{i) $a_{01}=\varepsilon_1$}

\begin{align}
&\widetilde{\boldsymbol{\Psi}}_\text{id} (a_{01} =\varepsilon_1,\boldsymbol{\varepsilon}, \Lambda, \mathbf{z}) \pm \widetilde{\boldsymbol{\Psi}}_{(01)} (a_{01} =\varepsilon_1,\boldsymbol{\varepsilon}, \Lambda, \mathbf{z}) \nonumber \\
&= e^{\frac{1}{\varepsilon_2} \left( \pm \frac{\Lambda^2}{\varepsilon_1} + \frac{\Lambda^4}{4 \varepsilon_1 ^3} \mp \frac{\Lambda^6}{12 \varepsilon_1 ^5} + \frac{\Lambda^8}{96 \varepsilon_1 ^7} \pm \frac{11 \Lambda^{10}}{720 \varepsilon_1 ^9} + \mathcal{O}(\Lambda^{12}) \right)} \nonumber \\ 
& \quad \left[ z^{1/2} \pm z^{-1/2} + \frac{\Lambda^2}{\varepsilon_1 ^2} \frac{z^{3/2} \pm z^{-3/2}}{2} + \frac{\Lambda^4}{\varepsilon_1 ^4} \left( \frac{z^{5/2} \pm z^{-5/2}}{12} - \frac{z^{1/2} \pm z^{-1/2}}{8} - \frac{z^{-3/2} \pm z^{3/2}}{12} \right) \right. \nonumber \\
&  + \frac{\Lambda^6}{\varepsilon_1 ^6} \left( \frac{z^{7/2} \pm z^{-7/2}}{144} + \frac{\mp z^{5/2} - z^{-5/2}}{18} - \frac{ z^{3/2} \pm z^{-3/2}}{48} +\frac{\pm z^{1/2} +z^{-1/2}}{8} \right)  \nonumber \\ 
& + \frac{\Lambda^8}{\varepsilon_1 ^8} \left( \frac{z^{9/2} \pm z^{-9/2}}{2880} +\frac{\mp z^{7/2} - z^{-7/2}}{192} + \frac{49(\pm z^{3/2} + z^{-3/2})}{28} - \frac{37(z^{1/2} \pm z^{-1/2})}{1152} \right) \nonumber \\
&+ \frac{\Lambda^{10}}{\varepsilon_1 ^{10}} \left( \frac{z^{11/2} \pm z^{-11/2}}{86400} +\frac{\mp z^{9/2} -z^{-9/2}}{3600} + \frac{z^{7/2} \pm z^{-7/2}}{5760} \right. \nonumber \\
& \left. \left. \quad \quad\quad + \frac{41(\pm z^{5/2} + z^{-5/2})}{1152} - \frac{317(z^{3/2} \pm z^{-3/2})}{2304} - \frac{121(\pm z^{1/2} + z^{-1/2})}{1728} \right) + \mathcal{O}(\Lambda^{12}) + \mathcal{O}(\varepsilon_2) \right]
\end{align}
Using the dictionary \eqref{energydeg} we compute
\begin{align}
E_{2, m=1} ^\pm = \frac{\varepsilon_1 ^2}{8} \mp \frac{\Lambda^2}{2} - \frac{\Lambda^4}{4 \varepsilon_1 ^2} \pm \frac{\Lambda^6}{8 \varepsilon_1 ^4} - \frac{\Lambda^8}{48 \varepsilon_1 ^6} \mp \frac{11 \Lambda^{10}}{288 \varepsilon_1 ^8} + \mathcal{O}(\Lambda^{12}).
\end{align}
These split eigenvalues and split eigenstate wavefunctions exactly match with the known results for the Mathieu function. 

\paragraph{ii) $a_{01}=2\varepsilon_1$}
\begin{align}
&\widetilde{\boldsymbol{\Psi}}_\text{id} (a_{01} =2\varepsilon_1,\boldsymbol{\varepsilon}, \Lambda, \mathbf{z}) \pm \widetilde{\boldsymbol{\Psi}}_{(01)} (a_{01} =2\varepsilon_1,\boldsymbol{\varepsilon}, \Lambda, \mathbf{z}) \nonumber \\
&= e^{\frac{1}{\varepsilon_2} \left( \left(-\frac{1}{3} \mp \frac{1}{2} \right)\frac{\Lambda^4}{\varepsilon_1 ^3} + \left( \frac{379}{864} \pm \frac{4}{9} \right) \frac{\Lambda^8}{\varepsilon_1 ^7} + \mathcal{O}(\Lambda^{12}) \right) } \nonumber \\
& \left[ z \pm z^{-1} + \frac{\Lambda^2}{\varepsilon_1 ^2} \left( \frac{z^2 \pm z^{-2}}{3} - (1 \pm 1) \right) + \frac{\Lambda^4}{\varepsilon_1 ^4} \left( \frac{z^3 \pm z^{-3}}{24} - \left( \frac{5}{9} \pm \frac{1}{2} \right)(z \pm z^{-1})  \right) \right. \nonumber \\
&+ \frac{\Lambda^6}{\varepsilon_1 ^6} \left( \frac{z^4 \pm z^{-4}}{360} + \left( -\frac{7}{72} \mp \frac{1}{18} \right)(z^2 \pm z^{-2}) + \frac{49}{18} (1 \pm 1) \right) \nonumber \\
&+ \frac{\Lambda^8}{\varepsilon_1 ^8} \left( \frac{z^5 \pm z^{-5}}{8640} - \left( \frac{1}{120} \pm \frac{1}{576} \right)(z^3 \pm z^{-3}) + \left( \frac{25655}{10368} \pm \frac{133}{54} \right)(z\pm z^{-1}) \right) \nonumber \\
&+ \frac{\Lambda^{10}}{\varepsilon_1 ^{10}} \left( \frac{z^6 \pm z^{-6}}{302400} - \left( \frac{11}{25920} \mp \frac{1}{14400}  \right) (z^4 \pm z^{-4}) + \left( \frac{91283}{155520} \pm \frac{37}{64} \right) (z^2 \pm z^{-2}) -\frac{134855}{5184} (1\pm 1) \right) \nonumber \\
&\left. + \mathcal{O}(\Lambda^{12}) + \mathcal{O}(\varepsilon_2) \right]
\end{align}

\begin{align}
E_{2,m=2} ^\pm = \frac{\varepsilon_1 ^2}{2} + \left( \frac{1}{3} \pm \frac{1}{2} \right) \frac{\Lambda^4}{\varepsilon_1 ^2} - \left( \frac{379}{432} \pm \frac{8}{9} \right) \frac{\Lambda^8}{\varepsilon_1 ^6} + \mathcal{O}(\Lambda^{12})
\end{align}

\paragraph{iii) $a_{01}=3 \varepsilon_1 $}
\begin{align}
&\widetilde{\boldsymbol{\Psi}}_\text{id} (a_{01} =3\varepsilon_1,\boldsymbol{\varepsilon}, \Lambda, \mathbf{z}) \pm \widetilde{\boldsymbol{\Psi}}_{(01)} (a_{01} =3\varepsilon_1,\boldsymbol{\varepsilon}, \Lambda, \mathbf{z}) \nonumber \\
&= e^{\frac{1}{\varepsilon_2} \left( - \frac{\Lambda^4}{8 \varepsilon_1 ^3} \pm \frac{\Lambda^6}{12 \varepsilon_1 ^5} - \frac{13\Lambda^8}{1280 \varepsilon_1 ^7} \mp \frac{\Lambda^{10}}{64 \varepsilon_1 ^9} + \mathcal{O}(\Lambda^{12}) \right)} \nonumber \\
& \left[ z^{3/2} \mp z^{-3/2} + \frac{\Lambda^2}{\varepsilon_1 ^2} \left( \frac{z^{5/2} \mp z^{-5/2}}{4} - \frac{z^{1/2} \mp z^{-1/2}}{2} \right) + \frac{\Lambda^4}{\varepsilon_1 ^4} \left( \frac{z^{7/2} \mp z^{-7/2}}{40} - \frac{5(z^{3/2} \mp z^{-3/2})}{32} - \frac{z^{1/2} \mp z^{-1/2}}{4} \right) \right. \nonumber \\
&+ \frac{\Lambda^6}{\varepsilon_1 ^6} \left( \frac{z^{9/2} \mp z^{-9/2}}{720} - \frac{11(z^{5/2} \mp z^{-5/2})}{640} + \frac{\mp z^{3/2} + z^{-3/2}}{8} +\frac{z^{1/2} \mp z^{-1/2}}{64} \right) \nonumber \\
& + \frac{\Lambda^8}{\varepsilon_1 ^8} \left( \frac{z^{11/2} \mp z^{-11/2}}{20160} - \frac{11(z^{7/2} \mp z^{-7/2})}{11520} + \frac{\mp z^{5/2} + z^{-5/2}}{64} - \frac{1621(z^{3/2} \mp z^{-3/2})}{51200} + \frac{21(\pm z^{1/2} - z^{-1/2})}{128} \right) \nonumber \\
&+ \frac{\Lambda^{10}}{\varepsilon_1 ^{10}} \left( \frac{z^{13/2} \mp z^{-13/2}}{806400} - \frac{z^{9/2} \pm z^{-9/2}}{32256} +\frac{3( \mp z^{7/2} + z^{-7/2})}{3200} -\frac{12329(z^{5/2} \pm z^{-5/2})}{1843200} \right. \nonumber \\
& \left. \left. \quad \quad \quad \quad+ \frac{9(\pm z^{3/2} - z^{-3/2})}{128} + \frac{14061(z^{1/2} \mp z^{-1/2})}{102400}  \right) + \mathcal{O}(\Lambda^{12}) + \mathcal{O}(\varepsilon_2) \right]
\end{align}
\begin{align}
E_{2, m=3} ^\pm = \frac{9 \varepsilon_1^2}{8} + \frac{\Lambda^4}{8 \varepsilon_1 ^2} \mp \frac{\Lambda^6}{8 \varepsilon_1 ^4} + \frac{13 \Lambda^8}{640 \varepsilon_1 ^6} \pm \frac{5 \Lambda^{10}}{128 \varepsilon_1 ^8} + \mathcal{O}(\Lambda^{12})
\end{align}

\subsection{$N=3$} \label{examcomp2}
In the case of $N=3$ we do not have a known result to compare to. Although the degenerate perturbative expansion can be done in principle for the non-Hermitian Hamiltonians, it quickly becomes tedious for increasing orders. The following results from gauge theory provides an alternative way to compute the split eigenfunctions and the split eigenvalues.
\paragraph{i) $a_{01}=a_{02}= \varepsilon_1$}
\tiny
\begin{align}
[&\widetilde{\boldsymbol{\Psi}}_\text{(012)} (\boldsymbol{a} ,\boldsymbol{\varepsilon}, \Lambda, \mathbf{z}) + \zeta \widetilde{\boldsymbol{\Psi}}_\text{(021)} (\boldsymbol{a} ,\boldsymbol{\varepsilon}, \Lambda, \mathbf{z}) + \zeta^2 \widetilde{\boldsymbol{\Psi}}_\text{id} (\boldsymbol{a},\boldsymbol{\varepsilon}, \Lambda, \mathbf{z}) ]\vert_{a_{01} = a_{02} = \varepsilon_1} \nonumber \\
&= e^{\frac{1}{\varepsilon_2} \left( \zeta \frac{\Lambda^2}{\varepsilon_1} + \zeta^2 \frac{\Lambda^4}{2 \varepsilon_1 ^3} + \frac{\Lambda^6}{12 \varepsilon_1 ^5} - \zeta \frac{3 \Lambda^8}{9 \varepsilon_1 ^7} + \mathcal{O}(\Lambda^{10}) \right) } z_0 ^{-\frac{a_0}{\varepsilon_1}} z_1 ^{-\frac{a_0}{\varepsilon_1} +1} z_2 ^{-\frac{a_0}{\varepsilon_1} +1} \nonumber \\
&\left[ 1  + \zeta \frac{z_0}{z_1} + \zeta^2 \frac{z_0}{z_2} + \frac{\Lambda^2}{\varepsilon_1 ^2} \left(  \frac{z_1}{2 z_0} + \frac{z_2}{z_1}  + \zeta \left(  \frac{z_0 z_2}{2 z_1 ^2} + \frac{z_0 ^2}{z_1 z_2}  \right) +\zeta^2 \left( \frac{z_0 ^2}{2 z_2 ^2} +\frac{z_1}{z_2} \right)  \right) \right. \nonumber \\
&+ \frac{\Lambda^4}{\varepsilon_1 ^4} \left( - \frac{z_0}{2z_1} - \frac{z_1 ^2}{12 z_0 ^2} - \frac{z_0 ^2}{4 z_2 ^2} - \frac{z_1}{2 z_2} + \zeta \left( \frac{3 z_0 ^2}{4 z_1 ^2} - \frac{z_1}{4 z_0} + \frac{z_0 ^3}{4 z_1 z_2 ^2} - \frac{z_0}{2 z_2} \right)  + \zeta^2 \left( - \frac{1}{2} + \frac{z_0 ^3}{12 z_2 ^3} + \frac{3 z_0 z_1}{4 z_2^2} - \frac{z_0 ^2}{z_1 z_2} + \frac{z_1 ^2}{4 z_0 z_2} \right) \right) \nonumber \\
& + \frac{\Lambda^6}{\varepsilon_1 ^6} \left( \frac{1}{8} + \frac{z_1 ^3}{144 z_0 ^3} - \frac{z_0 ^3}{z_2 ^3} -\frac{3 z_0 z_1}{4 z_2 ^2} - \frac{5 z_0 ^2}{4 z_1 z_2} - \frac{z_1 ^2}{6 z_0 z_2} - \frac{z_0 z_2}{4 z_1 ^2} + \frac{z_1 z_2}{6 z_0 ^2} + \frac{z_2 ^2}{4 z_0 z_1} + \frac{z_2 ^3}{36 z_1 ^3} \right. \nonumber \\
&\left. \quad \quad + \zeta \left( \frac{z_0}{8 z_1} - \frac{z_1 ^2}{18 z_0 ^2} + \frac{z_0 ^4}{36 z_1 z_2 ^3} - \frac{z_0 ^2}{4 z_2 ^2} + \frac{z_0 ^3}{4 z_1 ^2 z_2} - \frac{5 z_1}{4 z_2} - \frac{5 z_1}{4 z_2} - \frac{3 z_2}{4 z_0} + \frac{z_0 ^2 z_2 }{6 z_1 ^3} - \frac{z_2 ^2}{6 z_1 ^2} + \frac{z_0 z_2 ^3}{144 z_1 ^4} \right) \right. \nonumber \\
&\left. \quad \quad + \zeta^2 \left( - \frac{3 z_0 ^2}{4 z_1 ^2} - \frac{z_1}{4 z_0} + \frac{z_0 ^4}{144 z_2 ^4} + \frac{z_0 ^2 z_1}{6 z_2 ^3} - \frac{z_0 ^3}{6 z_1 z_2 ^2} + \frac{z_1 ^2}{4 z_2 ^2} + \frac{z_0}{8 z_2 ^2} + \frac{z_1 ^3}{36 z_0 ^2 z_2} - \frac{5 z_2}{4 z_1} - \frac{z_0 z_2 ^2}{18 z_1 ^3} \right) \right) \nonumber \\
&+ \frac{\Lambda^8}{\varepsilon_1 ^8} \left( -\frac{3 z_0 ^2} {4 z_1 ^2} + \frac{z_1}{4 z_0} + \frac{ z_1 ^4}{2880 z_0 ^4} - \frac{z_0 ^4}{192 z_2 ^4} -\frac{7 z_0 ^2 z_1}{36 z_2 ^3} - \frac{25 z_0 ^3}{72 z_1 z_2 ^2} - \frac{7 z_1 ^2}{24 z_2 ^2} + \frac{z_1 ^3}{48 z_0 ^2 z_1} + \frac{13 z_2}{8 z_1} +\frac{5 z_1 ^2 z_2}{288 z_0 ^3} + \frac{5 z_2 ^2}{72 z_0 ^2} - \frac{z_0 z_2 ^2}{24 z_1 ^3} + \frac{5 z_2 ^3}{144 z_0 z_1 ^2} + \frac{z_2 ^4}{576 z_1 ^4} \right. \nonumber \\
& \quad \quad + \zeta \left( \frac{25}{16} + \frac{5 z_0 ^3}{72 z_1 ^3} - \frac{z_1 ^3}{192 z_0 ^3} + \frac{z_0 ^5}{576 z_1 z_2 ^4} - \frac{z_0 ^3}{24 z_2 ^3} + \frac{5 z_0 ^4}{144 z_1 ^2 z_2 ^2} - \frac{3 z_0 z_1}{4 z_2 ^2} + \frac{13 z_0 ^2}{8 z_1 z_2} - \frac{25 z_1 ^2}{72 z_0 z_2} + \frac{z_0 z_2}{4 z_1 ^2} - \frac{7 z_1 z_2}{36 z_0 ^2} + \frac{5 z_0 ^2 z_2 ^2}{288 z_1 ^4} - \frac{7 z_2 ^2}{24 z_0 z_1} - \frac{ z_2 ^3}{48 z_1 ^3}+ \frac{z_0 z_2 ^4}{2880 z_1 ^5} \right) \nonumber \\
& \left. \quad \quad + \zeta^2 \left( + \frac{25 z_0}{16 z_1} - \frac{z_1 ^2}{24 z_0 ^2} + \frac{z_0 ^5}{2880 z_2 ^5} + \frac{5 z_0 ^3 z_1}{288 z_2 ^4} - \frac{z_0 ^4}{48 z_1 z_2 ^3} + \frac{5 z_0 z_1 ^2}{72 z_2 ^3} + \frac{z_0 ^2}{4 z_2 ^2} + \frac{5 z_1 ^3}{144 z_0 z_2 ^2} - \frac{ 7 z_0 ^3}{24 z_1 ^2 z_2} + \frac{13 z_1}{8 z_2} + \frac{ z_1 ^4}{576 z_0 ^3 z_2}- \frac{3 z_2}{4 z_0} - \frac{ 7 z_0 ^2 z_2 }{36 z_1 ^3} - \frac{25 z_2 ^2}{72 z_1 ^2}- \frac{z_0 z_2 ^3}{192 z_1 ^4} \right) \right) \nonumber \\
&\left. + \mathcal{O}(\Lambda^{10}) + \mathcal{O}(\varepsilon_2) \right]
\end{align}
\normalsize
From the dictionary \eqref{dics} we compute
\begin{align}
E_{2,m=1} ^\zeta &= \frac{\varepsilon_1 ^2}{3} - \zeta \Lambda^2 - \zeta^2 \frac{\Lambda^4}{\varepsilon_1 ^2} - \frac{\Lambda^6}{4\varepsilon_1 ^4} + \zeta \frac{3 \Lambda^8}{2 \varepsilon_1 ^6} + \mathcal{O}(\Lambda^{10}) \\
E_{3, m=1} ^\zeta &= \frac{2 \varepsilon_1 ^3}{27} + \zeta \Lambda^2 (-2 a_0 + \varepsilon_1 ) - \zeta^2 \frac{2(a_0 -\varepsilon_1) \Lambda^4}{\varepsilon_1 ^2} - \frac{(2a_0 + \varepsilon_1)\Lambda^6}{4 \varepsilon_1 ^4} + \zeta \frac{3(4a_0 - 3 \varepsilon_1) \Lambda^8}{4 \varepsilon_1 ^6} + \mathcal{O}(\Lambda^{10}).
\end{align}

\paragraph{ii) $a_{01} = a_{02} = 2 \varepsilon_1$}
\tiny
\begin{align}
[&\widetilde{\boldsymbol{\Psi}}_\text{(012)} (\boldsymbol{a} ,\boldsymbol{\varepsilon}, \Lambda, \mathbf{z}) + \zeta \widetilde{\boldsymbol{\Psi}}_\text{(021)} (\boldsymbol{a} ,\boldsymbol{\varepsilon}, \Lambda, \mathbf{z}) + \zeta^2 \widetilde{\boldsymbol{\Psi}}_\text{id} (\boldsymbol{a},\boldsymbol{\varepsilon}, \Lambda, \mathbf{z}) ]\vert_{a_{01} = a_{02} = 2\varepsilon_1} \nonumber \\
&= e^{\frac{1}{\varepsilon_2} \left( -\zeta \frac{\Lambda^4}{2 \varepsilon_1 ^3} - \frac{2 \Lambda^6}{27 \varepsilon_1 ^5} + \zeta^2 \frac{5 \Lambda^8}{16 \varepsilon_1 ^7} + \mathcal{O}(\Lambda^{10}) \right)} z_0 ^{-\frac{a_0}{\varepsilon_1}} z_1 ^{-\frac{a_0}{\varepsilon_1} +2} z_2 ^{-\frac{a_0}{\varepsilon_1} +2} \nonumber \\
&\left[ 1+\zeta\frac{z_0 ^2}{z_1 ^2} + \zeta^2 \frac{z_0 ^2}{z_2 ^2} + \frac{\Lambda^2}{\varepsilon_1 ^2} \left( \frac{z_1}{3 z_0} - \frac{z_0}{z_2} + \frac{z_2}{z_1} + \zeta \left( - \frac{z_0}{z_1} +\frac{z_0 ^3}{z_1 ^2 z_2} + \frac{z_0 ^2 z_2}{z_1 ^3} \right) + \zeta^2 \left( \frac{z_0 ^3}{3 z_2 ^3} + \frac{z_0 z_1}{z_2 ^2} - \frac{z_0 ^2}{z_1 z_2} \right) \right) \right. \nonumber \\
&+ \frac{\Lambda^4}{\varepsilon_1 ^4} \left( \frac{z_1 ^2}{24 z_0 ^2} - \frac{z_0 ^2}{2 z_2 ^2} - \frac{2 z_1}{z_2} + \frac{4 z_2}{9z_0} + \frac{z_2 ^2}{4 z_1 ^2} +\zeta \left( - \frac{1}{2} + \frac{4 z_0 ^3}{9 z_1 ^3} + \frac{z_0 ^4}{4 z_1 ^2 z_2 ^2} - \frac{2 z_0 z_2}{3 z_1 ^2} + \frac{z_0 ^2 z_2 ^2}{24 z_1 ^4} \right) + \zeta^2 \left( -\frac{z_0 ^2}{2 z_1 ^2} + \frac{z_0 ^4}{24 z_2 ^4} + \frac{4 z_0 ^2 z_1}{9 z_2 ^3} - \frac{2 z_0 ^3}{3 z_1 z_2 ^2} + \frac{z_1 ^2}{4 z_2 ^2} \right) \right) \nonumber \\
&+ \frac{\Lambda^6}{\varepsilon_1 ^6} \left( \frac{14}{27} + \frac{z_1 ^3}{360z_0 ^3} - \frac{z_0 ^3}{18 z_2 ^3} - \frac{z_0 z_1}{6 z_2 ^2} + \frac{3 z_0 ^2}{2 z_1 z_2} + \frac{3 z_0 ^2}{2 z_1 z_2} - \frac{z_1 ^2}{8 z_0 z_2} + \frac{z_0 z_2}{4 z_1 ^2} + \frac{5 z_1 z_2}{72 z_0 ^2} + \frac{5 z_2 ^2}{36 z_0 z_1} + \frac{z_2 ^3}{36 z_1 ^3} \right. \nonumber \\
& \quad\quad + \zeta \left( \frac{14 z_0 ^2}{27 z_1 ^2} - \frac{z_1}{18 z_0} + \frac{z_0 ^5}{36 z_1 ^2 z_2 ^3} + \frac{z_0 ^3}{4 z_1 z_2 ^2} + \frac{3 z_0}{2z_2} + \frac{5 z_0 ^4}{36 z_1 ^3 z_2} + \frac{5 z_0 ^3 z_2}{72 z_1 ^4} - \frac{z_2}{6 z_1} - \frac{z_0 z_2 ^2}{8 z_1 ^3} + \frac{z_0 ^2 z_2 ^3}{360 z_1 ^5} \right) \nonumber \\
&\left. \quad\quad + \zeta^2 \left( \frac{3 z_0}{2 z_1} + \frac{z_0 ^5}{360 z_2 ^5} + \frac{5 z_0 ^3 z_1}{72 z_2 ^4} - \frac{z_0 ^4}{8 z_1 z_2 ^3} + \frac{5 z_0 z_1 ^2}{36 z_2 ^3} + \frac{14 z_0 ^2}{27 z_2 ^2} + \frac{z_1 ^3}{36 z_0 z_2 ^2} - \frac{z_0 ^3}{6 z_1 ^2 z_2} + \frac{z_1}{4z_2} - \frac{z_0 ^2 z_2}{18 z_1 ^3} + \frac{z_0 z_2 ^2}{8 z_1 ^3} \right) \right) \nonumber \\
& + \frac{\Lambda^8}{\varepsilon_1 ^8} \left( \frac{35 z_0 ^2}{32 z_1 ^2} + \frac{29 z_1}{162 z_0} + \frac{z_1 ^4}{8640z_0 ^4} - \frac{z_0 ^4}{576z_2 ^4} + \frac{2z_0 ^2 z_1}{27 z_2 ^3} + \frac{7 z_0 ^3}{9 z_1 z_2 ^2} - \frac{z_1 ^2}{96 z_2 ^2} - \frac{44 z_0}{27 z_2} - \frac{z_1^3}{90 z_0 ^2 z_2} + \frac{61 z_2}{54 z_1}+ \frac{z_1 ^2 z_2}{180 z_0 ^3} + \frac{5 z_2 ^2}{192 z_0 ^2} + \frac{z_0 z_2 ^2}{18 z_1 ^3} + \frac{z_2 ^3}{54 z_0 z_1 ^2} + \frac{z_2 ^4}{576 z_1 ^4} \right. \nonumber \\
&\quad\quad + \zeta \left( \frac{5 z_0 ^4}{192 z_1 ^4} - \frac{44 z_0}{27 z_1} - \frac{z_1 ^2}{576z_0 ^2} + \frac{ z_0 ^6}{576 z_1 ^2 z_2 ^4} + \frac{z_0 ^4}{18 z_1 z_2 ^3} + \frac{35 z_0 ^2}{32 z_2 ^2} + \frac{z_0 ^5}{54 z_1 ^3 z_2 ^2} + \frac{61 z_0 ^3}{54 z_1 ^2 z_2} + \frac{7 z_1}{9 z_2} + \frac{2 z_2}{27 z_0} + \frac{29 z_0 ^2 z_2}{162 z_1 ^3} + \frac{z_0 ^3 z_2 ^2}{180 z_1 ^5} - \frac{z_2 ^2}{96 z_1 ^2} - \frac{z_0 z_2 ^3}{90 z_1 ^4} + \frac{z_0 ^2 z_2 ^4}{8640 z_1 ^6}  \right) \nonumber \\
&\left.\quad\quad + \zeta^2 \left( \frac{35}{32} + \frac{2 z_0 ^3}{27 z_1 ^3} + \frac{z_0 ^6}{8640z_2 ^6} + \frac{z_0 ^4 z_1}{180 z_2 ^5} - \frac{z_0 ^5}{90 z_1 z_2 ^4} + \frac{z_0 ^2 z_1 ^2}{192 z_2 ^4} + \frac{29 z_0 ^3}{162 z_2 ^3} + \frac{z_1 ^3}{54 z_2 ^3} - \frac{z_0 ^4}{96 z_1 ^2 z_2 ^2} + \frac{61 z_0 z_1}{54 z_2 ^2} + \frac{z_1 ^4}{576 z_0 ^2 z_2 ^2} - \frac{44 z_0 ^2}{27 z_1 z_2} + \frac{z_1 ^2}{18 z_0 z_2} + \frac{7 z_0 z_2}{9 z_1 ^2} - \frac{z_0 ^2 z_2 ^2}{576 z_1 ^4} \right) \right) \nonumber \\
&\left. + \mathcal{O}(\Lambda^{10}) + \mathcal{O}(\varepsilon_2) \right]
\end{align}
\normalsize
\begin{align}
E_{2, m=2} ^\zeta &= \frac{4 \varepsilon_1 ^2}{3} + \zeta \frac{\Lambda^4}{\varepsilon_1 ^2} + \frac{2 \Lambda^6}{9 \varepsilon_1 ^4} - \zeta^2 \frac{5 \Lambda^8}{4 \varepsilon_1 ^6} + \mathcal{O}(\Lambda^{10}) \\
E_{3, m=2} ^\zeta &= \frac{16 \varepsilon_1 ^3}{27} +\zeta \frac{2(a_0 -\varepsilon_1) \Lambda^4}{\varepsilon_1 ^2} + \frac{4(a_0 - 4\varepsilon_1) \Lambda^6}{9 \varepsilon_1 ^4} - \zeta^2 \frac{(5 a_0 -6 \varepsilon_1 )\Lambda^8}{2 \varepsilon_1 ^6} + \mathcal{O}(\Lambda^{10})
\end{align}

\begin{thebibliography}{99}

\bibitem{sw1}
N. Seiberg, E. Witten, \emph{Monopole Condensation, And Confinement In N=2 Supersymmetric Yang-Mills Theory}, \emph{Nucl.Phys.} {\bf B426} (1994) 19-52. [\href{https://arxiv.org/abs/hep-th/9407087}{hep-th/9407087}]

\bibitem{sw2}
N. Seiberg, E. Witten, \emph{Monopoles, Duality and Chiral Symmetry Breaking in N=2 Supersymmetric QCD}, \emph{Nucl.Phys.} {\bf B431} (1994) 484-550. [\href{https://arxiv.org/abs/hep-th/9408099}{hep-th/9408099}]

\bibitem{swint1}
A. Gorsky, I. Krichever, A. Marshakov, A. Mironov, A. Morozov, \emph{Integrability and Seiberg-Witten Exact Solution}, \emph{Phys.Lett.} {\bf B355} (1995) 466-474. [\href{https://arxiv.org/abs/hep-th/9505035}{hep-th/9505035}]

\bibitem{swint2}
R. Donagi, E. Witten, \emph{Supersymmetric Yang-Mills Systems And Integrable Systems}, \emph{Nucl.Phys.} {\bf B460} (1996) 299-334. [\href{https://arxiv.org/abs/hep-th/9510101}{hep-th/9510101}]

\bibitem{swint3}
E. Martinec, N. Warner, \emph{Integrable systems and supersymmetric gauge theory}, \emph{Nucl.Phys.} {\bf B459} (1996) 97-112. [\href{https://arxiv.org/abs/hep-th/9509161}{hep-th/9509161}]

\bibitem{nek1}
N. Nekrasov, \emph{Seiberg-Witten Prepotential From Instanton Counting}, \emph{Adv.Theor.Math.Phys.} {\bf 7} (2004) 831-864. [\href{https://arxiv.org/abs/hep-th/0206161}{hep-th/0206161}]

\bibitem{nekokoun}
N. Nekrasov, A. Okounkov, \emph{Seiberg-Witten Theory and Random Partitions}, [\href{https://arxiv.org/abs/hep-th/0306238}{hep-th/0306238}]

\bibitem{nek5}
A. Losev, A. Marshakov, N. Nekrasov, \emph{Small Instantons, Little Strings and Free Fermions}, [\href{https://arxiv.org/abs/hep-th/0302191v3}{hep-th/0302191v3}]

\bibitem{nek6}
A. Marshakov, N. Nekrasov, \emph{Extended Seiberg-Witten Theory and Integrable Hierarchy}, \emph{JHEP} {\bf 0701} (2007) 104. [\href{https://arxiv.org/abs/hep-th/0612019v2}{hep-th/0612019v2}]

\bibitem{neksha1}
N. Nekrasov, S. Shatashvili, \emph{Supersymmetric vacua and Bethe ansatz}, \emph{Nucl.Phys.Proc.Suppl.} {\bf 192-193} (2009) 91-112. [\href{https://arxiv.org/abs/0901.4744}{arXiv:0901.4744}]

\bibitem{neksha2}
N. Nekrasov, S. Shatashvili, \emph{Quantum integrability and supersymmetric vacua}, \emph{Prog.Theor.Phys.Suppl.} {\bf 177} (2009) 105-119. [\href{https://arxiv.org/abs/0901.4748}{arXiv:0901.4748}]

\bibitem{neksha3}
N. Nekrasov, S. Shatashvili, \emph{Quantization of Integrable Systems and Four Dimensional Gauge Theories}, [\href{https://arxiv.org/abs/0908.4052}{arXiv:0908.4052}]

\bibitem{nek9}
N. Nekrasov, \emph{Five Dimensional Gauge Theories and Relativistic Integrable Systems}, \emph{Nucl.Phys.B.} {\bf 531} (1998) 323-344. [\href{https://arxiv.org/abs/hep-th/9609219}{hep-th/9609219}]

\bibitem{nekwit}
N. Nekrasov, E. Witten, \emph{The Omega Deformation, Branes, Integrability, and Liouville Theory}, \emph{JHEP} {\bf 09} (2010) 092. [\href{https://arxiv.org/abs/1002.0888v2}{arXiv:1002.0888v2}]

\bibitem{gukwit}
S. Gukov, E. Witten \emph{Branes and Quantization}, \emph{ATMP} {\bf 13} (2009) 1445-1518. [\href{https://arxiv.org/abs/0809.0305v2}{arXiv:0809.0305v2}]

\bibitem{kantachi}
H. Kanno, Y. Tachikawa, \emph{Instanton counting with a surface operator and the chain-saw quiver}, \emph{JHEP} {\bf 06} (2011) 119 [\href{https://arxiv.org/abs/1105.0357}{arXiv:1105.0357}]

\bibitem{nekpes}
N. Nekrasov, V. Pestun, \emph{Seiberg-Witten geometry of four dimensional N=2 quiver gauge theories}, [\href{https://arxiv.org/abs/1211.2240}{arXiv:1211.2240}]

\bibitem{nekpessha}
N. Nekrasov, V. Pestun, S. Shatashvili, \emph{Quantum geometry and quiver gauge theories}, [\href{https://arxiv.org/abs/1312.6689}{arXiv:1312.6689}]

\bibitem{nek2}
N. Nekrasov, \emph{BPS/CFT correspondence: non-perturbative Dyson-Schwinger equations and $qq$-characters}, \emph{JHEP} {\bf 03} (2016) 181 [\href{https://arxiv.org/abs/1512.05388}{arXiv:1512.05388}]

\bibitem{nek3}
N. Nekrasov, \emph{BPS/CFT correspondence II: Instantons at crossroads, Moduli and Compactness Theorem}, \emph{Adv.Theor.Math.Phys.} {\bf 21} (2017) 503-583 [\href{https://arxiv.org/abs/1608.07272}{arXiv:1608.07272}]

\bibitem{nek4}
N. Nekrasov, \emph{BPS/CFT Correspondence III: Gauge Origami partition function and $qq$-characters}, \emph{Comm.Math.Phys.} {\bf 358} (2018) 863-894 [\href{https://arxiv.org/abs/1701.00189}{arXiv:1701.00189}]

\bibitem{nek7}
N. Nekrasov, \emph{BPS/CFT Correspondence IV: sigma models and defects in gauge theory}, \emph{Lett.Math.Phys.} (2018) 1-44 [\href{https://arxiv.org/abs/1711.11011}{arXiv:1701.11011}]

\bibitem{nek8}
N. Nekrasov, \emph{BPS/CFT Correspondence V: BPZ and KZ equations from $qq$-characters}, [\href{https://arxiv.org/abs/1711.11582}{arXiv:1701.11582}]

\bibitem{gms}
A. Gorsky, A. Milekhin, N. Sopenko, \emph{Bands and gaps in Nekrasov partition function}, \emph{JHEP} {\bf 01} (2018) 133 [\href{https://arxiv.org/abs/1712.02936}{arXiv:1712.02936}]

\bibitem{bra}
A. Braverman, \emph{Instanton counting via affine Lie algebras I: Equivariant J-functions of (affine) flag manifolds and Whittaker vectors}, [\href{https://arxiv.org/abs/math/0401409}{math/0401409}]

\bibitem{braeti}
A. Braverman, P. Etingof, \emph{Instanton counting via affine Lie algebras II: from Whittaker vectors to the Seiberg-Witten prepotential}, [\href{https://arxiv.org/abs/math/0409441}{math/0409441}]

\bibitem{gai1}
D. Gaiotto, \emph{N=2 dualities}, \emph{JHEP} {\bf 08} (2012) 034 [\href{https://arxiv.org/abs/0904.2715}{arXiv:0904.2715}]

\bibitem{koztes}
K. K. Kozlowski, J. Teschner, \emph{TBA for the Toda chain}, [\href{https://arxiv.org/abs/1006.2906v1}{arXiv:1006.2906v1}]

\bibitem{hit1}
N. Hitchin, \emph{Stable Bundles and Integrable Systems}, \emph{Duke Math. J.} {\bf 54} (1987) 91-114.

\bibitem{gut}
M. C. Gutzwiller, \emph{The Quantum Mechanical Toda Lattice}, \emph{Ann. Phys.} {\bf 124} (1980) 347-381.

\bibitem{gp}
M. Gaudin, V. Pasquier, \emph{The Periodic Toda Chain and a Matrix Generalization of the Bessel Function Recursion Relations}, \emph{J.Phys. A: Math. Gen} {\bf 25} (1992) 5243-5252.

\bibitem{kl1}
S. Kharchev, D. Lebedev, \emph{Integral Representation for the Eigenfunctions of a Quantum
Periodic Toda Chain}, \emph{Lett.Math.Phys.} {\bf 50} (1999) 53-77.

\bibitem{kl2}
S. Kharchev, D. Lebedev, \emph{Integral Representations for the Eigenfunctions of Quantum Open and Periodic Toda Chains from QISM Formalism}, \emph{J.Phys.} {\bf A34} (2001) 2247-2258.

\bibitem{an}
D. An, \emph{Complete set of Eigenfunctions of the quantum Toda chain}, \emph{Lett.Math.Phys.} {\bf 87} (2009) 209-223.

\bibitem{agt}
L. F. Alday, D. Gaiotto, Y. Tachikawa, \emph{Liouville Correlation Functions from Four-dimensional Gauge Theories}, \emph{Lett.Math.Phys.} {\bf 91} (2010) 167-197. [\href{https://arxiv.org/abs/0906.3219v2}{arXiv:0906.3219v2}]

\bibitem{agtw}
N. Wyllard, \emph{$A_{N-1}$ conformal Toda field theory correlation functions from conformal N=2 SU(N) quiver gauge theories}, \emph{JHEP} {\bf 0911:002} (2009). [\href{https://arxiv.org/abs/0907.2189v2}{arXiv:0907.2189v2}]

\bibitem{agt2}
L. F. Alday, D. Gaiotto, Y. Tachikawa, H. Verlinde, \emph{Loop and surface operators in N=2 gauge theory and Liouville modular geometry}, \emph{JHEP} {\bf 01} (2010) 113. [\href{https://arxiv.org/abs/0909.0945v3}{arXiv:0909.0945v3}]

\bibitem{gai2}
D. Gaiotto, \emph{Surface Operators in N=2 4d Gauge Theories}, \emph{JHEP} {\bf 11} (2012) 90. [\href{http://xxx.lanl.gov/abs/0911.1316}{arXiv:0911.1316}]

\bibitem{tes}
N. Drukker, J. Gomis, T. Okuda, J. Teschner, \emph{Gauge Theory Loop Operators and Liouville Theory}, \emph{JHEP} {\bf 1002:057} (2010). [\href{https://arxiv.org/abs/0909.1105v3}{arXiv:0909.1105v3}]

\bibitem{pesoku}
J. Gomis, T. Okuda, V. Pestun, \emph{Exact Results for 't Hooft Loops in Gauge Theories on $S^4$}, \emph{JHEP} {\bf 05} (2012). [\href{https://arxiv.org/abs/1105.2568v1}{arXiv:1105.2568v1}]

\bibitem{frgutes}
E. Frenkel, S. Gukov, J. Teschner, \emph{Surface Operators and Separation of Variables}, \emph{JHEP} {\bf 01} (2016) 179. [\href{https://arxiv.org/abs/1506.07508}{arXiv:1506.07508}]

\bibitem{dunun}
G. V. Dunne, M. Unsal, \emph{Generating Non-perturbative Physics from Perturbation Theory}, \emph{Phys. Rev. D} {\bf 89} (2014). [\href{https://arxiv.org/abs/1306.4405}{arXiv:1306.4405}] \\
G. V. Dunne, M. Unsal, \emph{Uniform WKB, Multi-instantons, and Resurgent Trans-Series}, \emph{Phys. Rev. D} {\bf 89} (2014). [\href{https://arxiv.org/abs/1401.5202}{arXiv:1401.5202}]

\bibitem{kre}
D. Krefl, \emph{Non-Perturbative Quantum Geometry II}, \emph{JHEP} {\bf 12} (2014) 118. [\href{https://arxiv.org/abs/1410.7116}{arXiv:1410.7116}]

\bibitem{bkk}
M. Bullimore, H-C. Kim, P. Koroteev, \emph{Defects and Quantum Seiberg-Witten Geometry}, \emph{JHEP} {\bf 05} (2015) 95. [\href{https://arxiv.org/abs/1412.6081}{arXiv:1412.6081}]

\bibitem{gokdun}
G. Basar, G. V. Dunne, \emph{Resurgence and the Nekrasov-Shatashvili Limit: Connecting Weak and Strong Coupling in the Mathieu and Lam'e Systems}, \emph{JHEP} {\bf 1502} (2015) 160. [\href{https://arxiv.org/abs/1501.05671}{arXiv:1501.05671}] \\
G. V. Dunne, M. Unsal, \emph{WKB and Resurgence in the Mathieu Equation}, [\href{https://arxiv.org/abs/1603.04924}{arXiv:1603.04924}]

\bibitem{algen}
J-E. Bourgine, M. Fukuda, K. Harada, Y. Matsuo, R-D. Zhu, \emph{(p,q)-webs of DIM representations, 5d N=1 instanton partition functions and qq-characters}, \emph{JHEP} {\bf 11} (2017) 34. [\href{https://arxiv.org/abs/1703.10759}{arXiv:1703.10759}] \\
H. Awata, H. Kanno, A. Mironov, A. Morozov, K. Suetake, Y. Zenkevich, \emph{(q,t)-KZ equations for quantum toroidal algebra and Nekrasov partition functions on ALE spaces}, \emph{JHEP} {\bf 03} (2018) 192. [\href{https://arxiv.org/abs/1712.08016}{arXiv:1712.08016}]

\bibitem{wzh}
A. Grassi, Y. Hatsuda, M. Marino, \emph{Topological Strings from Quantum Mechanics}, [\href{https://arxiv.org/abs/1410.3382}{arXiv:1410.3382}] \\
X. Wang, G. Zhang, M. Huang, \emph{New Exact Quantization Condition for Toric Calabi-Yau Geometries}, \emph{Phys. Rev. Lett.} {\bf 115} (2015). [\href{https://arxiv.org/abs/1505.05360}{arXiv:1505.05360}] \\
S. Franco, Y. Hatsuda, M. Marino, \emph{Exact quantization conditions for cluster integrable systems}, [\href{https://arxiv.org/abs/1512.03061}{arXiv:1512.03061}] \\
D. Krefl, \emph{Non-Perturbative Quantum Geometry III}, \emph{JHEP} {\bf 08} (2016). [\href{https://arxiv.org/abs/1605.00182}{arXiv:1605.00182}]

\bibitem{Cor}
L. Cornalba, \emph{D-brane Physics and Noncommutative Yang-Mills Theory}, \emph{Adv.Theor.Math.Phys.} {\bf 4} (2000) 271-281. [\href{https://arxiv.org/abs/hep-th/9909081}{hep-th/9909081}] \\
N. Ishibashi, \emph{A Relation between Commutative and Noncommutative Descriptions of D-branes}, [\href{https://arxiv.org/abs/hep-th/9909176}{hep-th/9909176}]

\bibitem{bpz}
A. A. Belavin, A. M. Polyakov, A. B. Zamolodchikov, \emph{Infinite conformal symmetry in two-dimensional quantum field theory}, \emph{Nucl.Phys.B} {\bf 241} (1984) 333-380.

\bibitem{wang}
Z. X. Wang, D. R. Guo, \emph{Special Functions}, \emph{World Scientific} (1989).





\end{thebibliography}


\end{document}